\documentclass[a4paper,11pt]{article}
\pdfoutput=1 

\usepackage{jcappub} 

\usepackage[T1]{fontenc} 
\usepackage{cleveref}
\usepackage{graphicx}
\usepackage{caption}
\usepackage{subcaption}
\usepackage{ulem} 
\usepackage{bm}

\def\bx{{\bf x}}
\def\bk{{\bf k}}

\def\bp{{\bf p}}

\title{Stability in Axion Inflation with Strong Backreaction from a Massive Vector Boson}

\author{Michael J.~Baker, Joaquim Iguaz Juan and Lorenzo Sorbo}

\affiliation{Amherst Center for Fundamental Interactions, Department of Physics, University of Massachusetts Amherst, MA 01003, USA}

\emailAdd{mjbaker@umass.edu}
\emailAdd{jiguazjuan@umass.edu}
\emailAdd{sorbo@umass.edu}

\abstract{We study a modification of the model of axion inflation coupled to a $U(1)$ gauge field where the vector field is {\it massive}. In the conventional scenario with a massless gauge field, the onset of the regime where the gauge field strongly backreacts on the inflaton displays an instability whose nonlinear evolution and endpoint remain poorly understood. We argue that, if the gauge field is massive enough, the transition to the strong backreaction regime instead occurs smoothly, avoiding this instability. This observation suggests that axion inflation with massive gauge fields admits a controllable strong backreaction regime, so that a phenomenologically viable realization of inflation might be possible in this class of models.}

\begin{document}
\maketitle
\flushbottom

\section{Introduction}
\label{sec:introduction}

In a few years primordial inflation will be  half a century old~\cite{Starobinsky:1980te,Guth:1980zm,Sato:1981qmu,Linde:1981mu,Albrecht:1982wi}.  During these decades, the impressive agreement between the predictions of simple inflationary scenarios and increasingly precise cosmological observations has established inflation as the standard framework for describing the earliest moments of observable cosmic history. At the same time, however, inflation is a paradigm, not a specific theory. A wide variety of models have been proposed, differing in field content, interactions and underlying theoretical motivations. Among these, a strong theoretical motivation and a rich phenomenology have made {\it axion inflation} a particularly fertile framework for exploring connections between early-Universe cosmology, particle physics and quantum gravity.  

On the more formal side, a guiding principle for the construction of inflationary models is the requirement that the inflaton potential remain sufficiently flat over the relevant field range. Generic scalar field theories receive radiative corrections that tend to spoil this flatness. A particularly attractive way of satisfying this requirement is to assume an approximate shift symmetry for the inflaton~\cite{Freese:1990rb}. Axions are especially compelling realizations of this idea. Moreover, axion-like fields arise ubiquitously in ultraviolet completions such as string theory, making them well-motivated inflaton candidates from both bottom-up and top-down perspectives. 

Phenomenologically, axion inflatons are naturally coupled to gauge fields\footnote{While these gauge fields can belong to any (Abelian or non-Abelian) gauge group, in this work we will focus on couplings to a single vector degree of freedom originating from a $U(1)$ gauge group.} via a Chern-Simons coupling. In the presence of  this coupling, the rolling inflaton excites the quanta of the vector field, with one of its two helicities achieving exponentially large occupation numbers~\cite{Anber:2006xt}. This population of quanta leads to multiple phenomenological predictions, including nongaussianities~\cite{Barnaby:2010vf}, deviations from scale invariance~\cite{Namba:2015gja}, formation of a population of primordial black holes~\cite{Linde:2012bt}, generation of primordial chiral gravitational waves at CMB~\cite{Sorbo:2011rz} or interferometer~\cite{Cook:2011hg} frequencies, baryogenesis~\cite{Anber:2015yca}, as well as the possible generation of cosmologically relevant magnetic fields~\cite{Garretson:1992vt}. See~\cite{Pajer:2013fsa} for a review.

\subsection{The Strong Backreaction Regime of Axion Inflation}

The occupation number of vectors in axion inflation depends exponentially on a quantity proportional to $|\dot\Phi|/H$, the axion velocity divided by the Hubble parameter, which typically increases during inflation. As a consequence, axion inflation typically ends in a regime where backreaction of vectors on the inflaton cannot be neglected~\cite{Adshead:2015pva}. Importantly, much of the interesting phenomenology of axion inflation occur precisely during the final stages of inflation, when backreaction effects are not negligible.

A second reason to study the strong backreaction regime of axion inflation is the following. In ordinary slow roll inflation with potential $V(\Phi)$ the slow roll parameters,
\begin{align}\label{eq:def_sr}
    \epsilon\equiv \frac{1}{2}\left(\frac{M_{\rm Pl}}{V(\Phi)}\,\frac{d\,V(\Phi)}{d\Phi}\right)^2\qquad {\rm and}\qquad \eta\equiv\frac{M_{\rm Pl}}{V(\Phi)}\,\frac{d^2V(\Phi)}{d\Phi^2}\,,
\end{align}
must be much smaller than unity in absolute value throughout the region of  potential $V(\Phi)$ in which inflation occurs.  As mentioned above, however, flatness of the inflationary potential is difficult to achieve.\footnote{Even the simplest models of axion inflation, such as the original model of natural inflation \cite{Freese:1990rb}, are conjectured~\cite{Banks:2003sx,Arkani-Hamed:2006emk} to have a potential that is too steep to support inflation when embedded in a UV-complete theory.}  Production of vectors drains energy from the rolling zero mode of the inflaton, which in the strong backreaction regime leads to a sizable slowing down of the inflaton~\cite{Anber:2009ua}.  This mechanism is analogous to that operating in the context of warm inflation~\cite{Berera:1995ie} and trapped inflation~\cite{Green:2009ds}. As a consequence, a steady regime of controlled strong backreaction of the produced vectors on the rolling inflaton could lead to slow-roll inflation even for potentials with $\epsilon$ and/or $|\eta|$ larger than unity, opening up new avenues in inflationary model building.

Unfortunately, despite a significant amount of work done over the last decade or so, the strong backreaction regime in axion inflation coupled to $U(1)$ gauge fields is still not fully understood. Work, including refs.~\cite{Cheng:2015oqa, Notari:2016npn, Sobol:2019xls, DallAgata:2019yrr, Domcke:2020zez, Garcia-Bellido:2023ser, vonEckardstein:2023gwk}, that used various numerical techniques has shown that, if one imposes that the inflaton remains spatially homogeneous, the strong backreaction regime is characterized by wild and apparently chaotic oscillations of the inflaton velocity $\dot\Phi$. Allowing for a spatially inhomogeneous inflaton leads to a very different picture~\cite{Figueroa:2023oxc,Figueroa:2024rkr}: after one or two oscillations in $\dot\Phi$, large inflaton gradients develop (although note that this outcome is not universal~\cite{Barbon:2025wjl}), leading to a new steady state with large spatial perturbations but much less dramatic time dependence. Unfortunately, the realistic case of inhomogeneous inflaton could so far only be studied with lattice computations~\cite{Caravano:2022epk, Figueroa:2023oxc, Figueroa:2024rkr, Sharma:2024nfu, Iarygina:2025ncl} (see also~\cite{Domcke:2023tnn}), which can only cover a limited range ($\lesssim10$ e-folds for the most recent works) due to excessive computational costs. Ideally, one would track the evolution of spatial inhomogeneities over the whole inflationary period to better gauge their relevance.

Why do oscillations in $\dot\Phi$ emerge during the transition to the strong backreaction regime?  Neglecting the spatial gradients of the inflaton, the equations of motion for the coupled inflaton-vector system can be written schematically (for the explicit expressions, see \cref{eq:eom-2.4,eq:eom-2.5} with $m_A=0$) as
\begin{align}\label{eq:eom_schematic1}
    &\ddot\Phi+3\,H\,\dot\Phi+\frac{d\,V(\Phi)}{d\Phi} =\langle {\cal O}_1 (A_\mu)\rangle\,,\\
    \label{eq:eom_schematic2}
    &{\cal O}_2[H,\,\dot\Phi]\,A_\mu=0\,,
\end{align}
where $\Phi$ is the (classical) inflaton field, $A_\mu$ is the (quantum) gauge field, and ${\cal O}_1$ and ${\cal O}_2$ are relatively simple operators (with ${\cal O}_2$ depending on the Hubble parameter and the inflaton velocity). 

The first attempt at solving this system without neglecting $\langle {\cal O}_1 (A_\mu)\rangle$ was made in ref.~\cite{Anber:2009ua}, whose authors used the fact that \cref{eq:eom_schematic2} can be solved exactly for constant $\dot\Phi$ and $H$ and then plugged the solution for $A_\mu$ into \cref{eq:eom_schematic1}, thus obtaining a differential equation for $\Phi$ only.  Note that this solution is not consistent, since $\dot\Phi$ is assumed to be constant when solving \cref{eq:eom_schematic2}, but is subsequently treated as time-dependent in \cref{eq:eom_schematic1}. Solutions of the equation obtained this way show a smooth transition between the weak and the strong backreaction regime of axion inflation (see for instance the red dashed lines on the right panels of fig.~6 in ref.~\cite{Domcke:2020zez}).

Unfortunately, as discussed above, numerical solution of the system of \cref{eq:eom_schematic1,eq:eom_schematic2} shows that this transition is not smooth at all (see the black solid lines on the right panels of fig.~6 in ref.~\cite{Domcke:2020zez}). As argued in refs.~\cite{Notari:2016npn,Domcke:2020zez}, the discrepancy between the solution provided in~\cite{Anber:2009ua} and those found in the subsequent numerical work originates from the fact that a change in $\dot\Phi$ is not immediately felt by all the modes in $A_\mu$, but only by those that are being amplified when the change occurs. As a consequence, the expectation value $\langle {\cal O}_1 (A_\mu)\rangle$ depends on the value of $\dot\Phi$ computed at an earlier time than that at which the left hand side of \cref{eq:eom_schematic1} is evaluated.  This delay is responsible for the oscillations in $\dot\Phi$ found in numerical solutions of \cref{eq:eom_schematic1,eq:eom_schematic2}. For this reason we will refer to the solution provided in ref.~\cite{Anber:2009ua} as the solution found in the {\it instant backreaction} (IBR) approximation.

\subsection{This Work}

The goal of this work is to propose a modification to the model of axion inflation that maintains much of its theoretical and  phenomenological successes while having a more controllable strong backreaction regime. Here, by ``more controllable'' we mean that the evolution in the strong backreaction regime follows the IBR solution. As argued in~\cite{Creminelli:2023aly}, one can expect this to be the case when the dynamics associated to particle production and the backreaction of the produced matter on the inflaton occurs on time- and length-scales much shorter than the Hubble scale.  Following this suggestion, we will explore the possibility that the degree of freedom coupled to inflaton is a {\it massive vector field}. We think of this as a Higgsed $U(1)$ gauge theory where the Higgs field is heavy and has decoupled. A mass larger than the Hubble parameter effectively decouples the intermediate and long wavelength modes of the vector, so that only short wavelength modes participate in the dynamics (which become effectively Minkowskian). We thus expect that sufficiently massive vectors will not experience the delayed backreaction responsible for the irregular behavior observed in axion inflation with a massless gauge field. A version of axion inflation coupled to massive {\it non-Abelian} gauge fields was discussed in~\cite{Adshead:2016omu}.

How do we show that massive vectors lead to IBR evolution in the strong backreaction regime? And how large must the vector mass be? Ref.~\cite{Peloso:2022ovc} provides an analytical study of the deviation of axion inflation with massless vectors from the IBR solution in the strong backreaction regime. That paper (and subsequently, using other techniques, refs.~\cite{vonEckardstein:2023gwk,Sobol:2026nfh}) shows that the equations of motion, linearized around the IBR solution, feature an unstable mode with complex Lyapunov exponent, corresponding to solutions that oscillate around the IBR solution with increasing amplitude, in agreement with  numerical results. 

In the present work we reproduce the analysis of~\cite{Peloso:2022ovc} using a semi-analytic approach, generalizing it to the case of a massive vector. We find that the mode functions of the vector for large vector mass $m_A$ are Boltzmann suppressed with respect to the de Sitter temperature $H/2\pi$ and depend exponentially on a quantity proportional to $(\alpha|\dot\Phi|/f-m_A)/H$, where $\alpha/f$ is the axion-vector coupling. For this reason, we vary $m_A/H$ in our stability analysis while keeping the quantity $(\alpha|\dot\Phi|/f-m_A)/H$ constant. We find that the real part of the largest Lyapunov exponent {\it decreases} in absolute value as we increase the ratio $m_A/H$. The instability disappears as this ratio exceeds a value that, for the part of parameter space we have explored, is of the order of $100$. We consider this evidence of the fact that axion inflation with heavy vectors ($m_A\gtrsim 100\,H$) evolves smoothly along the IBR solution even during the strong backreaction phase.

Our paper is organized as follows. In \cref{sec:model} we set up our Lagrangian and show that if the vector mass is larger than a few times the Hubble parameter the scalar component of the vector field can be neglected. In \cref{sec:solutions} we present the mode functions of the transverse components of the massive vector, estimate the magnitude of the backreaction term $\langle {\cal O}_1 (A_\mu)\rangle$ and present a rough estimate of the spectrum of scalar cosmological perturbations obtained in the strong backreaction regime.  This provides a formula which we use to estimate the values of certain parameters in the model. In \cref{sec:stability}, which contains the main results of this paper, we discuss the regimes of (in)stability of the IBR solution. Finally, we conclude and discuss some future directions in \cref{sec:conclusions}. Technical details can be found in three appendices.

\section{A Model of Axion Inflation with a Massive Vector Boson}
\label{sec:model}

We now discuss the model and derive the equations of motion for the physical degrees of freedom.  We will show that the longitudinal mode of the massive spin-one particle decouples and is not excited by the background dynamics.  Throughout this work we will use conformal time, $\tau =\int dt/a(t)$, where $t$ is cosmological time and $a(t)$ is the scale factor, and assume de Sitter expansion, so that $a(\tau) = -1/H\tau$ with a constant Hubble parameter $H$.  The metric on $(d\tau,dx,dy,dz)$ is then $g_{\mu\nu} = a^2(\tau)\,\text{diag}(-1,\, 1,\,1,\,1)$.

We consider a model with an axion inflaton, $\Phi$, and a massive spin-one particle, $A_\mu$, which we describe with a Proca action.  The Lagrangian density is
\begin{align}
    \label{eq:lagrangian}
    \mathcal{L}
    &=
    -\sqrt{-g}
    \left[
    \frac{1}{2}\partial_\mu \Phi \partial^\mu \Phi
    + 
    V(\Phi)
    +
    \frac{1}{4} F_{\mu\nu} F^{\mu\nu}
    + 
    \frac{\alpha}{4 f}\Phi F_{\mu\nu}\tilde{F}^{\mu\nu}
    + 
    \frac{m_A^2}{2} A_\mu A^\mu
    \right]
    \,,
\end{align}
where $\sqrt{-g} = a^4(\tau)$, 
\begin{align}
\label{eq:potential}
    V(\Phi) &= \frac{\Lambda^4}{2}
    \left(
    1+\cos\left(\frac{\Phi}{f}\right)
    \right)
    \,,
\end{align}
$\Lambda^4$ is the height of the potential, $f$ is the axion decay constant, $F_{\mu\nu} = \partial_\mu A_\nu - \partial_\nu A_\mu$, $\alpha$ is a dimensionless coupling parameter, $\tilde{F}_{\mu\nu} = \frac{1}{2\sqrt{-g}}\eta_{\mu\nu\alpha\beta}F^{\alpha\beta}$, $\eta_{\mu\nu\alpha\beta}$ is the four-dimensional Levi-Civita tensor in flat space with $\eta_{0123} = +1$ and $m_A$ is the mass of the spin-one field.  

We now separate into space and time, $A_\mu = (A_0,\,{\bf A})$ and $\partial_\mu=(\partial_\tau,\,{\bm\nabla})$, and write the three-dimensional vector field ${\bf A}$ as the gradient of a scalar field and a three-dimensional transverse vector field, ${\bf A}={\bm\nabla}\,A^s+{\bf A}^T$ with the condition ${\bm\nabla}\cdot\,{\bf A}^T=0$ (that is, we use a Helmholtz decomposition),
\begin{align}
    \label{eq:lagrangian-2}
    \mathcal{L}
    =&
    \frac{a^2}{2}
    \left(
    \partial_\tau \Phi \partial_\tau \Phi
    - {\bm\nabla}\Phi\cdot{\bm\nabla}\Phi
    \right)
    -
    a^4\, V(\Phi)
    - 
    \frac{\alpha}{f}\,\Phi\, \left(\partial_\tau {\bf A}^T\right)\cdot\left({\bm\nabla}\times{\bf A}^T\right)
    \notag
    \\
    &\,
    +\frac{1}{2}\left(\partial_\tau {\bf A}^T\right)\cdot\left(\partial_\tau {\bf A}^T\right)
    +
   \frac{1}{2}{\bf A}^T\cdot \nabla^2 {\bf A}^T
    - 
    \frac{a^2m_A^2}{2} \,{\bf A}^T\cdot{\bf A}^T
    \notag
    \\
    &\,
    - 
    \frac{1}{2}\partial_\tau A^s\,\nabla^2\,\partial_\tau A^s
    -
    \frac{1}{2}\,A_0\,\nabla^2A_0
    +
    \partial_\tau A^s \,\nabla^2 A_0
    +
    \frac{a^2m_A^2}{2} A_0^2
    + 
    \frac{a^2m_A^2}{2} A^s \,\nabla^2 A^s
    \,,
\end{align}
where we have dropped some total derivatives. We see here that $A_0$ and $A^s$ decouple from $\Phi$ (and from ${\bf A}^T$, as expected in a Helmholtz decomposition).  We also note that there are no time derivatives of $A_0$, which we will come to shortly.

From now on we will consider $\Phi$ as a classical field (that is, we actually consider $\langle\Phi\rangle$ but simply write it as $\Phi$), while we will keep treating ${\bf A}$  as a quantum field. Assuming spatial homogeneity in $\Phi$ (i.e., ${\bm\nabla}\Phi = 0$) and taking the mean-field Hartree approximation (which amounts to taking the vacuum expectation value) for the equation of motion of $\Phi$, the equations of motion are
\begin{align}
    \label{eq:eom-2.4}
    &\partial^2_\tau\Phi+2\frac{\partial_\tau a}{a}\,\partial_\tau\Phi+a^2\,
    \frac{d\, V(\Phi)}{d\Phi} 
    =
    -
    \frac{\alpha}{a^2\,f} 
    \langle 0|\left(\partial_\tau {\bf A}^T\right)\cdot\left({\bm\nabla}\times{\bf A}^T\right)
    |0\rangle\,,
    \\\label{eq:eom-2.5}
    &\left( \partial_\tau^2 - \nabla^2  +a^2 \,m_A^2   - \frac{\alpha}{f}\, (\partial_\tau \Phi)\, {\bm\nabla}\times \right)\,{\bf A}^T=0\,,
    \\
    \label{eq:eom-2.6}
    &\left(\partial_\tau^2 + a^2\,m_A^2\right) A^s =\partial_\tau A_0\,,
    \\
    \label{eq:eom-2.7}
    &\left(\nabla^2  -a^2\, m_A^2\right)A_0= \partial_\tau \nabla^2 A^s\,,
\end{align}
where, to obtain \cref{eq:eom-2.6}, we have used the fact that $\nabla^2 f(\bx) = 0$ implies $f(\bx)=0$ for any bounded function $f(\bx)$.  We now go to momentum space, writing
\begin{align}\label{eq:at_deco}
    {\bf A}^T 
    &=
    \sum_{\lambda=\pm}\int \frac{d^3{\bk}}{(2\pi)^{3/2}}
    \left[
    {\bm{\epsilon}}^\lambda({\bk})\,
    A_\lambda(\tau,\,k)\,
    \hat a_\lambda^T({\bk})
    e^{i\bk \cdot \bx}
    +
    \text{h.c.}
    \right]
    \\
    A^s
    &=
    \int \frac{d^3{\bk}}{(2\pi)^{3/2}}\,
    \hat A^s(\tau,\,\bk)\,
    e^{i{\bk} \cdot {\bx}}
    \\
    A_0
    &=
    \int \frac{d^3{\bk}}{(2\pi)^{3/2}}\,
    \hat A_0(\tau,\,\bk)\,
    e^{i{\bk} \cdot {\bx}}\,,
\end{align}
where $\bk$ is the comoving momentum,  $k = |\bk|$, $\hat a^T_\lambda(\bk)$ is an annihilation operator and ${\bm\epsilon}^\lambda(\bk)$ are circular polarisation vectors satisfying ${\bm\epsilon}^\lambda(\bk)\cdot{\bm\epsilon}^{\lambda'\ast}(\bk) = \delta_{\lambda \lambda'}$, $\bk\cdot{\bm\epsilon}^\lambda(\bk) = 0$ and $i\,{\bk}\times {\bm\epsilon}^\lambda({\bk}) = \lambda\,k\,{\bm\epsilon}^\lambda(\bk)$.  The mode functions for the transverse states are $A_\lambda(\tau,\,k)$ while $\hat A^s(\tau,\, \bk)$ and $\hat A_0(\tau,\,\bk)$ are Fourier transforms of the field operators.   We are treating the scalar modes differently to the transverse modes in anticipation of the next subsection.  Since $A^s$ is a real field we see that $(\hat A^s(\tau,\, \bk))^\dagger = \hat A^s(\tau,\,-\bk)$.  The equations of motion then become
\begin{align}
    \partial^2_\tau\Phi+2\frac{\partial_\tau a}{a}\,\partial_\tau\Phi
    + a^2
    \frac{d\, V(\Phi)}{d\Phi}=
    -
    \frac{\alpha}{a^2\,f} 
    \int &\frac{d^3\bk}{(2\pi)^3}\,
    \frac{k}{2}\,
    \partial_\tau\left(|A_+^2(\tau,\,k)| -\left|A_-^2(\tau,\,k)\right|\right)\,,\label{eq:kg_backr}
    \\
    \left(\partial_\tau^2 
    + 
    k^2 + a^2\,m_A^2
    - \lambda \frac{\alpha\,k}{f} (\partial_\tau \Phi) \right) A_\lambda(\tau,\,k)&=0\,,\label{eq:a_trans}
    \\
    \left(\partial_\tau^2 + a^2\,m_A^2\right) \hat A^s(\tau,\,\bk) &=\partial_\tau \hat A_0(\tau,\,\bk)\,, \label{eq:eom-A0}
    \\
    \left(k^2  +a^2\, m_A^2\right)\hat A_0(\tau,\,\bk)&= k^2\,\partial_\tau  \hat A^s(\tau,\,\bk)\,,
    \label{eq:eom-A0-mom-sp}
\end{align}
where we have used $[\hat{a}^T_\lambda(\bk),\,\hat{a}_{\lambda'}^{T\dagger}(\bk')] = \delta^{(3)}(\bk - \bk')\,\delta_{\lambda\lambda'}$ in the equation of motion of $\Phi$.  We saw in \cref{eq:lagrangian-2} that $\hat{A}_0$ has no time derivatives in the Lagrangian, so it is not a propagating field. In fact, \cref{eq:eom-A0-mom-sp} tells us that $A_0$ can be removed from the theory using
\begin{align}
\label{eq:A0-in-terms-of-As}
    \hat A_0(\tau,\,\bk)
    &=
    \frac{k^2}{k^2 + a^2\,m_A^2}\partial_\tau \hat A^s(\tau,\,\bk)
    \,.
\end{align}

\subsection{Adiabatic Evolution of the Longitudinal Mode}
\label{subsec:longitudinal}

We have noted in \cref{eq:lagrangian-2} that the longitudinal mode of the spin-one field decouples from the $\Phi$ field.  We will now show that the longitudinal mode evolves adiabatically for the parameters we are interested in, so that its quanta are not created during inflation.  We will first Fourier expand the fields in the Lagrangian, \cref{eq:lagrangian-2}, and substitute \cref{eq:A0-in-terms-of-As} into it.  Writing only the terms involving $A^s$ we find
\begin{align}
    L
    \supset&
    \int d^3\bk\,\frac{a^2\,m_A^2\,k^2}{k^2 + a^2\, m_A^2}\,
    \Bigg(
    \frac12\partial_\tau \hat A^s (\tau,\,\bk)
    \partial_\tau \hat A^{s\dagger} (\tau,\,\bk) 
    - \frac12\left(k^2 + a^2\,m_A^2\right) \hat A^s (\tau,\,\bk)
    \hat A^{s\dagger} (\tau,\,\bk)
    \Bigg)
    \,.
\end{align}
We now canonically normalize the field $A^s$ by defining
\begin{equation}
    \hat{A}^s_c(\tau,\,\bk)
    \equiv 
    z(\tau) \,\hat{A}^s(\tau,\,\bk) \,,
    \qquad\qquad z(\tau)\equiv\frac{a(\tau)\,m_A\,k}{\sqrt{k^2 + a^2(\tau)\,m_A^2}}
\,,
\end{equation}
which gives, after an integration by parts,
\begin{align}
    \mathcal{L}
    \supset
    \int d^3\bk
    \left[
    \frac12\partial_\tau \hat A^s_c(\tau,\,\bk)\,
    \partial_\tau \hat A^{s\dagger}_c(\tau,\,\bk)
    -\frac12
    \left(k^2+a^2\, m_A^2
    -\frac{\partial_\tau^2 z}{z}
    \right)
    \hat A^s_c(\tau,\,\bk) \,\hat A^{s\dagger}_c(\tau,\,\bk)
    \right]
    \,.
\end{align}
The equation of motion for the $\bk$-th mode is then 
\begin{align}
    \partial^2_\tau \hat A^s_c(\tau,\,\bk) + \left(k^2+ a^2\,m_A^2 -\frac{\partial_\tau^2 z}{z}\right)\,\hat A^s_c(\tau,\,\bk) = 0
    \,.
\end{align}

\begin{figure}
\centering
    \includegraphics[width=0.5\linewidth]{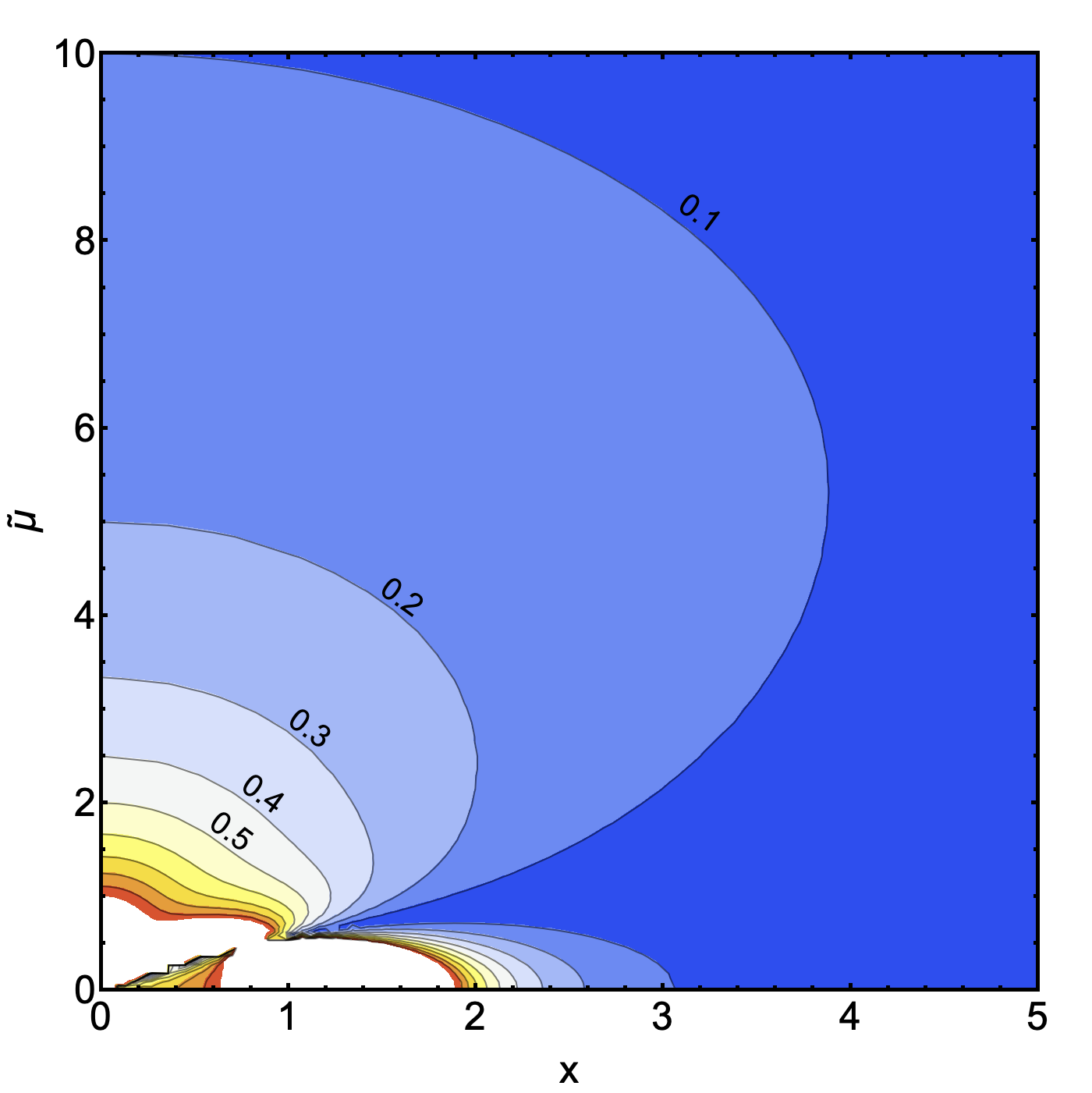}
    \caption{The function $|\partial_x\omega_s|/\omega_s^2$ shown as a function of $x$ and $\tilde\mu$.  The scalar degree of freedom $A_c^s$ evolves adiabatically if $|\partial_x\omega_s|/\omega_s^2 \ll 1$, while $|\partial_x\omega_s|/\omega_s^2 > 1$ in the white region.  Contour lines are shown with steps of 0.1 in the range 0 to 1.}
    \label{fig:adiabacity}
\end{figure}

Remembering that we assume de Sitter expansion, $a(\tau)=-1/H\tau$, and defining the ``time'' variable $x=-k\,\tau>0$, the equation of motion for $\hat{A}^s_c$ becomes
\begin{equation}
    \frac{d^2\hat A^s_c(x)}{dx^2}
    +
    \left(
    1+
    \frac{\tilde\mu^2}{x^2}+
    \frac{\tilde\mu^2-2\,x^2}{(x^2+\tilde\mu^2)^2}
    \right)\hat A^s_c(x)
    =0
    \,,
\end{equation}
where $\tilde\mu = m_A/H$.  We see that this is the equation of motion for a simple harmonic oscillator with a time dependent frequency,
\begin{equation}
    \omega_s= \sqrt{1+\frac{\tilde\mu^2}{x^2}+\frac{\tilde\mu^2 - 2\,x^2}{(x^2+\tilde\mu^2)^2}}
    \,.
\end{equation}

Particle production occurs when the frequency of the mode functions of the corresponding field evolve non-adiabatically.  That is, in the case of the scalar degree of freedom  $\hat A^s_c$, when $|\partial_x\omega_s|\gtrsim \omega_s^2$. We plot $|\partial_x\omega_s|/\omega_s^2$ in \cref{fig:adiabacity} and see numerically that production of quanta of $\hat A^s_c$ can be neglected for $\tilde\mu \gg 1$.  Algebraically, for $x=0$ we find that $\partial_x\omega_s/\omega_s^2 = -1/\tilde\mu$, for $x>0$ and $\tilde\mu^2>2$ the function $\partial_x\omega_s/\omega_s^2$ is negative definite and for $x>0$ and $\tilde\mu^2 > 5/3$ its derivative with respect to $x$, $\partial_x(\partial_x\omega_s/\omega_s^2)$, is positive.\footnote{These are sufficient conditions which we find by requiring all terms in relatively complicated expressions have the same signs.  Numerically, we find that $\partial_x\omega_s/\omega_s^2$ is negative and the derivative, $\partial_x(\partial_x\omega_s/\omega_s^2)$, is positive for a wider range of $\tilde\mu$.}  The absolute value $|\partial_x\omega_s|/\omega_s^2$ thus reduces monotonically as $x$ increases as long as $\tilde\mu ^2 > 2$, confirming that the scalar degree of freedom $\hat A^s_c$ can be neglected for $\tilde\mu \gg 1$.

\section{Solutions of the Equations of Motion for the Transverse Vector}
\label{sec:solutions}

Since the longitudinal mode decouples and evolves adiabatically, we only need to consider the equations of motion for $\Phi$ and the transverse component of the massive vector field, \cref{eq:kg_backr,eq:a_trans}.

The sign of the term $k^2 + a^2\,m_A^2 - \lambda \alpha\,k (\partial_\tau \Phi)/f$ in \cref{eq:a_trans} determines whether the mode oscillates or experiences exponential growth/decay.  Assuming $\partial_\tau\Phi > 0$, the term is always positive for $\lambda = -1$, so the $A_-$ mode simply oscillates.  However, if $\alpha \, k\, (\partial_\tau\Phi)/f$ is larger than $k^2 + a^2\,m_A^2$, then $A_+$ can experience exponential growth.  Neglecting the contribution from the $A_-$ modes and using spherical symmetry we can write the equation of motion for $\Phi$, \cref{eq:kg_backr}, as
\begin{align}
    \partial_\tau^2\Phi + 2\,\frac{\partial_\tau a}{a}\,\partial_\tau\Phi + a^2 \, \frac{dV(\Phi)}{d\Phi} 
    &= 
    - \frac{\alpha}{4 \pi^2\, a^2 \,f} \int dk \, k^3 \, \partial_\tau \left\vert A_+ \right\vert^2\,.
    \label{eq:KG_with_BR}
\end{align} 
In \cref{subsec:weak-back-reation} we will first solve the equation of motion \cref{eq:a_trans} for $A_+(\tau,\,k)$ on an inflating background under the approximation of constant inflaton velocity and Hubble parameter. In the subsequent \cref{subsec:strong-back-reaction} we will study the effect that the gauge modes obtained in \cref{subsec:weak-back-reation} have on the dynamics of the inflaton, through their contribution to the right hand side of \cref{eq:KG_with_BR}.

\subsection{Weak Backreaction Regime}
\label{subsec:weak-back-reation}

Assuming that $\partial_\tau\Phi/a\equiv \dot\Phi_0=$\,constant\,$>0$ and $H\equiv H_0=$\,constant, the equation of motion for $A_+$, \cref{eq:a_trans}, can be solved exactly in terms of Whittaker functions.  As in \cref{subsec:longitudinal}, we use the dimensionless time variable $x = -k\,\tau > 0$. We also introduce the two dimensionless parameters
\begin{align}\label{eq:xiandmu}
    \xi_0\equiv \alpha\,\frac{\dot\Phi_0}{2\,f\,H_0}>0\qquad {\rm and} \qquad \mu_0\equiv\sqrt{\frac{m_A^2}{H_0^2}-\frac14}\,.
\end{align}
The equation of motion for $A_+$ can then be written as 
\begin{align}\label{Apm} 
    \left(  \frac{d^2}{dx^2} + 1  -  \frac{2\, \xi_0 }{x} + \frac{\mu_0^2+1/4}{x ^2}   \right) A_+(x)
 = 0
 \,.
\end{align}
which, when written in terms of the variable $2ix$, is the well-known Whittaker equation. Except when we explicitly state otherwise, throughout the paper we will assume that the mass of the gauge field is much larger than the Hubble parameter, so $\mu_0^2+1/4\approx \mu_0^2$.

The solution for $A_+(x)$ that matches the standard Bunch-Davies vacuum, $e^{ix}/\sqrt{2k}$, at early times $k\,\tau \rightarrow - \infty$ (that is, for $x \gg 2\,\xi_0$ and $x \gg \sqrt{\mu_0^2+1/4}$) is
\begin{align} \label{Apmsol}
    A_{+}(k,x) =\frac{e^{\pi \xi_0/2 }}{\sqrt{2 k}} W_{i\xi_0, \,i\mu_0}(2i x)\equiv\frac{A_1(x)}{\sqrt{2k}}
    \,.
\end{align}
where $W$ denotes the Whittaker function.

\begin{figure}
\centering
\begin{subfigure}{.5\textwidth}
  \centering
  \includegraphics[width=1\linewidth]{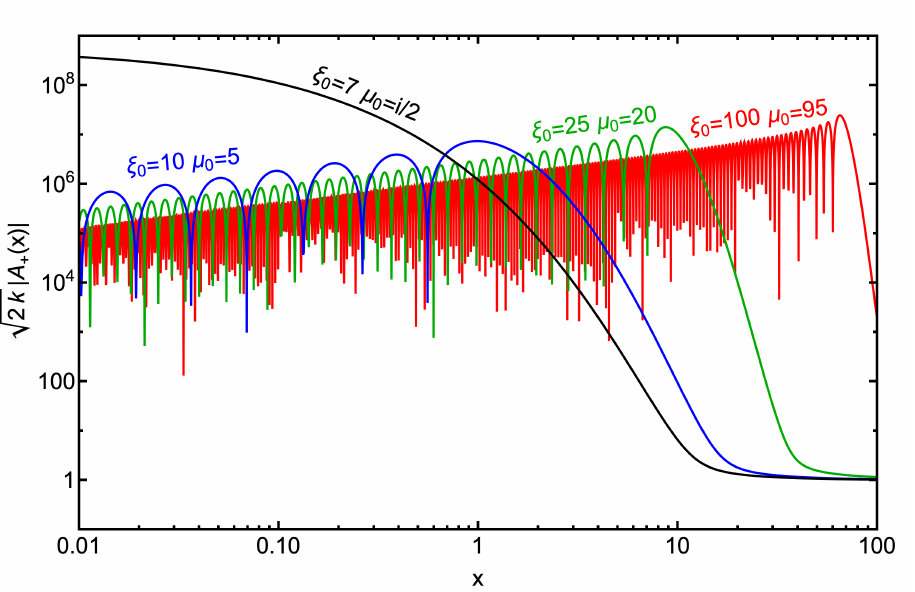}
  \label{fig:sub1}
\end{subfigure}%
\begin{subfigure}{.5\textwidth}
  \centering
  \includegraphics[width=1\linewidth]{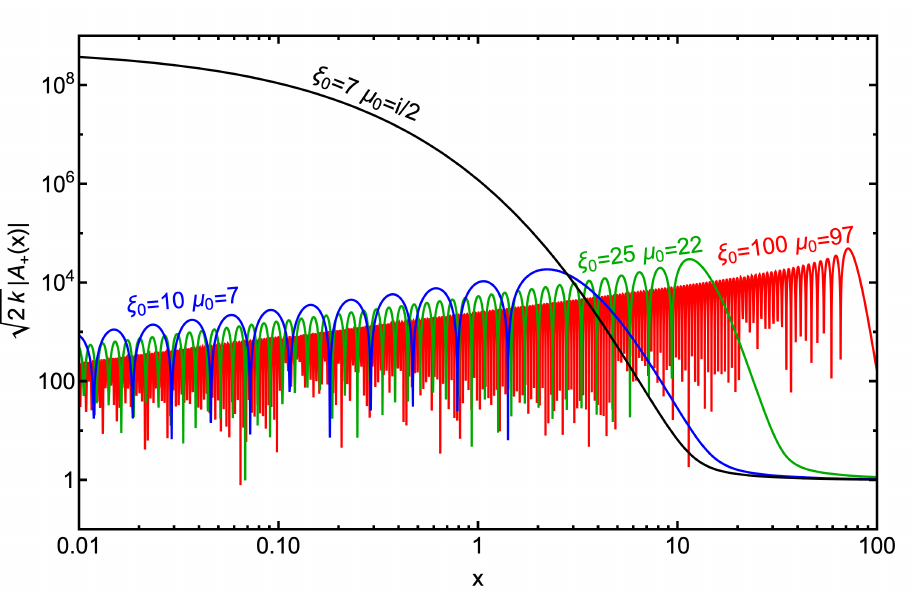}
  \label{fig:sub2}
\end{subfigure}
\caption{Norm of the mode function of the vector field, $|\sqrt{2k}\,A_+(x)|$, for $\xi_0-\mu_0=5$ (left) and $\xi_0-\mu_0=3$ (right). We also show the solution for the vector field in the massless case, $\mu_0=i/2$, for $\xi_0=7$ in black. Early physical times are given by $x\to\infty$, while $x\to 0$ corresponds to physical wavelengths much larger than the Hubble radius at a given time.}
\label{fig:Whittaker_function}
\end{figure}

In \cref{fig:Whittaker_function} we plot the absolute value of the function $A_1(x)$ for representative values of $\xi_0$ and $\mu_0$.  We see that for a massless vector boson ($\mu_0=i/2$) the absolute value of the mode functions is monotonically increasing as $x$ decreases and does not vanish in the limit $x\to 0$ (which corresponds to physical wavelengths much larger than the Hubble radius, $k/a(\tau) \ll H_0$).  For sufficiently massive vector bosons and $\xi_0-\mu_0\gtrsim {\cal O}(1)$, the amplitude of the mode functions is approximately proportional to $e^{\pi(\xi_0-\mu_0)}$ (see, e.g., \cref{eq:wkb_int} and \cref{eq:wkb_small} where we give expressions of the mode functions obtained using the WKB approximation). For this reason, in this paper we will mostly focus on the regime $\xi_0-\mu_0={\cal O}(1)>0$ with $\xi_0,\,\mu_0\gg 1$. In particular, in \cref{fig:Whittaker_function} we choose parameters with fixed values of $\xi_0-\mu_0=3,\,5$.  We see that the mode functions have a peak at $x\approx \xi_0$ and oscillate while decreasing as $x$ decreases.  The latter behavior can be understood as follows. At early times during inflation (i.e., for large values of $x$), the frequency in \cref{Apm} is real, the vector field modes are still in their vacuum, and the real and imaginary parts of the function $A_1(x)$ oscillate with amplitudes approximately equal to unity (with $|A_1(x)|$ remaining approximately constant). For intermediate values of $x$, then, \cref{Apm} acquires an imaginary frequency, leading to particle production, which is signaled by a tachyonic amplification of the mode functions. At even smaller values of $x$ the mode functions feel the effect of the mass term and oscillate back to $A_1(x)\to 0$. We denote the two transition points (i.e., the zeros of the frequency in \cref{Apm}) as $x_\pm$, with
\begin{align}\label{eq:xpm}
    x_\pm=\xi_0\pm\sqrt{\xi_0^2-\left(\mu_0^2+\frac14\right)}\,.
\end{align}
Since we will be interested in the regime $\xi_0\gg 1$ with $\xi_0-\mu_0={\cal O}(1)$, we will have $1\ll x_-\lesssim x_+\simeq \xi_0$.

\subsection{Strong Backreaction Regime}
\label{subsec:strong-back-reaction}

In this paper we are interested in the strong backreaction regime, when the right hand side of \cref{eq:KG_with_BR} cannot be neglected. In this subsection we estimate the integral appearing in that equation, assuming that both $\xi=\xi_0$ and $\mu=\mu_0$ are constant. In terms of the dimensionless time variable $x$, the integral reads
\begin{align}\label{eq:change_kx}
    \int dk \, k^3 \, \frac{\partial}{\partial \tau} \left\vert A_+(\tau,\,k) \right\vert^2= -\frac{1}{2\,\tau^4}\int dx \, x^3 \, \frac{d\,\left\vert A_1(x) \right\vert^2}{dx} \,.
\end{align}
Note that in principle the integral is divergent in the ultraviolet ($x\to\infty$), but we will cut it off at $x=x_+$ (see \cref{eq:xpm}), since for larger values of $x$ the transverse vector field modes are in their vacuum .

While the integral in \cref{eq:change_kx} can be computed numerically using the exact mode functions~(\ref{Apmsol}), it is useful to find an order of magnitude estimate which can be obtained using the WKB approximation of the mode functions (see \cref{app:wkb}).  We estimate the integral by noticing that it receives its main contribution from the region around $x\simeq x_- \approx \xi_0$, when the oscillatory behavior of the mode functions transitions to an exponential behavior. In Appendix~\ref{app:backreact} we obtain
\begin{align}\label{eq:back_int_approx}
\int dx \, x^3 \, \frac{d\,\left\vert A_1(x) \right\vert^2}{dx} 
&\approx -0.8\times e^{2\pi\,(\xi_0-\mu_0)}\frac{\xi_0^{7/2}}{(\xi_0-\mu_0)^{1/6}}\,,
\end{align}
where, as discussed above, we have assumed $1\lesssim \xi_0-\mu_0\ll \xi_0$.  We have checked numerically that this approximation is good at the $50\%$ level in the range $2\le \xi_0-\mu_0\le 10$ when $50<\xi_0<1000$.

By writing the equation of motion for $\Phi$, \cref{eq:KG_with_BR},  in physical time, setting $\partial_\tau\Phi/a=\dot\Phi_0=$\,constant\,$>0$ (which implies $d\,V(\Phi)/d\Phi<0$) and $H=H_0=$\,constant, and writing the resulting equation in terms of
$\xi_0$ and $\mu_0$, we obtain the background equation
\begin{align}\label{eq:back_const}
    \xi_0-\sqrt{\frac{\epsilon}{2}}\,\frac{\alpha\,M_P}{f}\simeq-\frac{\alpha^2\,H_0^2}{f^2}\,\frac{0.8}{48\pi^2}\times e^{2\pi\,(\xi_0-\mu_0)}\frac{\xi_0^{7/2}}{(\xi_0-\mu_0)^{1/6}}\,,
\end{align}
where we have assumed that the energy density during inflation is dominated by the potential, $3\,H_0^2\,M_P^2\simeq V(\Phi)$.  This is the {\it {``instant backreaction'' (IBR)}} solution.

In the weak backreaction regime the right hand side of \cref{eq:back_const} is negligible.  However, in the strong backreaction regime the term $\xi_0$ on the left hand side is subdominant with respect to the other two terms.

\subsection{Estimating the Spectrum of Scalar Perturbations in the Strong Backreaction Regime}
\label{subsec:Pspectrum}

Performing a rigorous calculation of the spectrum of perturbations in the strong backreaction regime is a challenging task. Here we follow the argument used in refs.~\cite{Anber:2009ua,Linde:2012bt} in the case of massless vector fields, expecting that it will give a good qualitative estimate of the spectrum of scalar perturbations in our case.  We will use this estimate to fix parameters in the subsequent discussion.

In our scenario, classical inhomogeneities in $\Phi$ are sourced by those in the gauge field. The curvature perturbation $\zeta(\bx)$ on uniform energy density hypersurfaces is given by the perturbation in the number of e-foldings, $\zeta(\bx)=\delta N(\bx)\equiv N(\bx)-\bar{N}$, where $\bar{N}$ is the number of e-foldings in the homogeneous background. If we write the perturbed value of the inflaton field as $\Phi=\Phi_{0}(\tau)+\delta\Phi(\tau,\,\bx)$, then 
\begin{align}\label{eq:defzeta}
  \zeta(\tau,\,\bx)=\frac{H}{\dot\Phi_0}\,\delta\Phi(\tau,\,\bx)\,.
\end{align}

In order to compute the power spectrum of $\zeta$, which observations show to be quasi scale-invariant and with an amplitude of the order of $10^{-9}$, we must therefore compute the correlators of $\delta\Phi$.  The fluctuation $\delta\Phi$ obeys the equation
\begin{align}\label{eq:kgpert}
\partial_\tau^2\delta\Phi+2\,\frac{\partial_\tau a}{a}\,\partial_\tau\delta\Phi+\left(-\nabla^2+a^2\frac{d^2\,V(\Phi)}{d\Phi^2}\right)\delta\Phi=&
-\frac{\alpha}{a^2\,f}\,\delta[{\bf E}\cdot{\bf B}]\,,
\end{align}
where we have introduced the notation
\begin{align}
    {\bf E}\cdot{\bf B}\equiv\partial_\tau {\bf A}^T\cdot({\bm\nabla}\times {\bf A}^T)
    \,,
\end{align}
The fluctuation term $\delta[{\bf E}\cdot{\bf B}]$ receives two contributions. First, it will contain intrinsic inhomogeneities, ${\bf E}\cdot{\bf B}-\langle{\bf E}\cdot{\bf B}\rangle$, that would exist even if the inflaton were to be spatially homogeneous. A second component comes from the fact that ${\bf E}\cdot{\bf B}$ depends on the value of $\Phi$ and $\dot\Phi$. We expect to obtain a reasonable approximation of the latter effect by identifying it with the dependence of the expectation value $\langle{\bf E}\cdot{\bf B}\rangle$ on $\Phi$, so that, if $\Phi$ is replaced by $\Phi+\delta\Phi$, then ${\bf E}\cdot{\bf B}$ receives a correction 
$\delta\Phi\, \partial \langle {\bf E}\cdot{\bf B}\rangle /\partial \Phi+\delta\dot\Phi\, \partial \langle {\bf E}\cdot{\bf B}\rangle /\partial \dot\Phi$. We therefore write
\begin{equation}\label{eq:rhspert}
\delta[{\bf E}\cdot{\bf B}]\simeq [{\bf E}\cdot{\bf B}-\langle {\bf E}\cdot{\bf B} \rangle]_{\delta\Phi=0}+\frac{\partial \langle{\bf E}\cdot{\bf B}\rangle}{\partial \Phi}\delta\Phi+\frac{\partial \langle{\bf E}\cdot{\bf B}\rangle}{\partial \dot\Phi}\delta\dot\Phi\,.
\end{equation}

To proceed we notice that the dependence of $\langle{\bf E}\cdot{\bf B}\rangle$ on $\Phi$ is slow-roll suppressed, so that the last two terms on the right hand side of eq.~(\ref{eq:rhspert}) are approximated by
\begin{align}
 \frac{\partial \langle{\bf E}\cdot{\bf B}\rangle}{\partial \Phi}\delta\Phi+\frac{\partial \langle{\bf E}\cdot{\bf B}\rangle}{\partial \dot\Phi}\delta\dot\Phi\simeq  \frac{\partial \langle{\bf E}\cdot{\bf B}\rangle}{\partial \dot\Phi}\delta\dot\Phi\simeq \pi\frac{\alpha\,\delta\dot\Phi}{f\,H}\langle{\bf E}\cdot{\bf B}\rangle\,. 
\end{align}
where we have also noticed that $\langle{\bf E}\cdot{\bf B}\rangle$ depends on $\dot\Phi$ mostly through its exponential dependence on $2\pi\,\xi_0$ (see, e.g., \cref{eq:back_int_approx}). We can thus write \cref{eq:kgpert}  as
\begin{align}\label{eq:kgpert-2}
\partial_\tau^2\delta\Phi+2\,\frac{\partial_\tau a}{a}\left(1+\frac{\pi}{2}\,\frac{\alpha^2 \langle{\bf E}\cdot{\bf B}\rangle}{a^4\,f^2\,H^2}\right)&\partial_\tau\delta\Phi+\left(-\nabla^2+a^2\frac{d^2\,V(\Phi)}{d\Phi^2}\right)\delta\Phi\nonumber\\
&=-\frac{\alpha}{a^2\,f}\,[{\bf E}\cdot{\bf B}-\langle {\bf E}\cdot{\bf B} \rangle]_{\delta\Phi=0}\,.
\end{align}

To estimate the magnitude of the various terms in this equation, we assume that the system is in the strong backreaction regime, which leads to slow roll inflation even if the inflationary potential would be too steep to support slow roll by itself. More specifically, we assume the inflationary potential to take the form given in \cref{eq:potential}, with $f\ll M_P$, while the potential energy of the inflaton dominates the energy in the Universe, so that $V(\Phi)\simeq 3\,H^2\,M_P^2$, $d\,V(\Phi)/d\Phi={\cal O}(H^2\,M_P^2/f)$ and $d^2\,V(\Phi)/d\Phi^2={\cal O}(H^2\,M_P^2/f^2)$. Furthermore, we assume that \cref{eq:kg_backr} simplifies in the strong backreaction regime to 
\begin{align}
    \langle{\bf E}\cdot{\bf B}\rangle\approx -a^4\,\frac{f}{\alpha}\,\frac{d\,V(\Phi)}{d\Phi}=a^4\,{\cal O}\left(\frac{H^2\,M_P^2}{\alpha}\right)\,.
\end{align}
Given these scalings, a study analogous to those of refs.~\cite{Anber:2009ua,Linde:2012bt} shows that the left hand side of \cref{eq:kgpert-2} is dominated by the portion of friction term proportional to $\langle{\bf E}\cdot{\bf B}\rangle$. Keeping only this term and using the definition \cref{eq:defzeta}, we obtain
\begin{align}\label{eq:zeta_int}
    \zeta(\tau,\,\bx)\simeq\frac{1}{2\pi\,\xi_0}\int^\tau\frac{d\tau'}{\tau'}\frac{[({\bf E}\cdot{\bf B})(\tau',\,\bx)-\langle {\bf E}\cdot{\bf B} \rangle(\tau')]_{\delta\Phi=0}}{\langle {\bf E}\cdot{\bf B} \rangle(\tau')}\,.
\end{align}

From this expression one can finally derive the power spectrum. As we discuss in \cref{app:powerspectrum}, numerical evaluation shows that the two point function of the curvature perturbation is proportional to $1/\xi_0^3$. Accounting for the numerical factors, we obtain 
\begin{align}\label{eq:pz_approx}
    {\cal P}_\zeta\approx \frac{0.05}{\xi_0^3}\,.
\end{align}
Equating this result to the observed value $ {\cal P}_\zeta\simeq 2\times 10^{-9}$ leads to the estimate $\xi_0\simeq 300$.

\section{Stability in the Strong Backreaction Regime}
\label{sec:stability}

In this section we present the main claim of our work: we argue that there is a region of the parameter space for the model (\ref{eq:lagrangian}) where the solution \cref{eq:back_const} is an attractor. That is, the evolution in the strong backreaction regime follows the {\it {``instant backreaction'' (IBR)}} solution.  To do so, we follow the philosophy of \cite{Peloso:2022ovc} and ask whether the solutions found in \cref{subsec:weak-back-reation,subsec:strong-back-reaction} are stable under small perturbations.  

We decompose the inflaton and positive helicity mode of the spin-one field into the background solution plus small perturbations,
\begin{align}
    \Phi(\tau) &= \bar{\Phi}(\tau)+\delta\Phi(\tau)
    \\
    A_+(\tau,\,k) &= \bar{A}_+(\tau,\,k) + \delta A_+(\tau,\,k),
\end{align}
where $\bar\Phi(\tau)$ is the IBR solution of \cref{eq:kg_backr,eq:a_trans}, i.e., it is obtained by solving \cref{eq:KG_with_BR} after replacing the integral on the right hand side with \cref{eq:change_kx,eq:back_int_approx}. Note that this solution is not rigorously correct for our system, since \cref{eq:back_int_approx} is obtained under the assumption of constant $\xi_0$ and $\mu_0$, which is not valid when deriving $\bar\Phi(\tau)$. The function $A_+(\tau,\,k)$ is the mode function of the positive-helicity transverse component of the vector field defined in \cref{eq:at_deco}, and $\bar{A}_+(\tau,\,k)$ is obtained under the same approximation of constant $\xi_0$ and $\mu_0$, so it is given by \cref{Apmsol}. To first order in perturbations, we obtain the equations 
\begin{align}
    &\partial_\tau^2\delta\Phi+2\frac{\partial_\tau a}{a}\,\partial_\tau\delta\Phi+a^2\frac{d^2\,V(\bar\Phi)}{d\bar\Phi^2}\delta\Phi
    =
    -\frac{\alpha}{f\, a^2}\int \frac{d^3\bk}{(2\pi)^3}\,\frac{k}{2}\,\frac{\partial}{\partial\tau} \left[\bar{A}_+\,\delta A^*_++\bar{A}^*_+\,\delta A_+ \right], 
    \label{eq:inf_eq}
    \\
    &\left(\partial_\tau^2+ k^2+m^2a^2-\frac{\alpha}{f}\,k\,\partial_\tau\bar{\Phi}\right)\,\delta A_+=
     \frac{\alpha}{f}\,k\,\bar{A}_+\,\partial_\tau\delta\Phi
     \,.
     \label{eq:gauge_eq}
\end{align}

Note that we do not perform a full perturbative study since we disregard metric perturbations and spatial inhomogeneities of the inflaton. These assumptions are motivated by the fact that we only want to evaluate the possible {\it onset} of an instability in the evolution of the zero mode of the inflaton. While lattice studies show that large inflaton gradients are generated after the instability develops~\cite{Figueroa:2023oxc,Figueroa:2024rkr}, these gradients are small at the onset of the instability so we expect their effect to be negligible.

If the IBR solution is unstable, then the instability develops quite quickly (see, e.g., fig.~6 in ref.~\cite{Domcke:2020zez} or fig.~2 in ref.~\cite{Garcia-Bellido:2023ser}) and the evolution of the background should be negligible. We will thus consider $\partial_t\bar\Phi=\dot\Phi_0$, $H=H_0$ and $d^2V(\bar\Phi)/d\bar\Phi^2$ as constants in this section.

We proceed analogously to the study of the linearized system of perturbations in ref.~\cite{Peloso:2022ovc}. We first formally solve the equation for the gauge field perturbation, \cref{eq:gauge_eq}, as a functional of the inflaton derivative $\partial_\tau\delta\Phi$ via the Green's function method. The Green's function satisfies the equation
\begin{equation}
    \left[ \partial_\tau^2+k^2+m^2a^2-\frac{\alpha\, k }{f} \,\partial_\tau\bar{\Phi}\right] G_k (\tau,\tau') = \delta(\tau-\tau').
\end{equation}
If $\bar{A}_+(\tau,\,k)$ is the solution (given in \cref{Apmsol}) of the associated homogeneous equation, \cref{Apm}, and $\bar{A}_+^*(\tau,\,k)$ is its complex conjugate, which is also a solution, then the retarded Green's function reads
\begin{align}\label{eq:green2}
    G_k(\tau,\tau') = -\frac{1}{k}\,\text{Im}\left[ \bar{A}_1(-k\tau)\, \bar{A}_1^*(-k\tau') \right]\,\theta(\tau-\tau'),
\end{align}
where we used the Wronskian $(\partial_{\tau'}\bar{A}_+(\tau',\,k))\,\bar{A}_+^*(\tau',\,k)-\bar{A}_+(\tau',\,k)\,(\partial_{\tau'}\bar{A}_+^{*}(\tau',\,k))=i$. Causality is guaranteed by the use of the {\it retarded} Green's function, which includes the Heaviside step function $\theta(\tau-\tau')$. The solution for the gauge field perturbation $\delta A$ can then be written as
\begin{equation}
    \delta A_+(\tau,\,k) = \frac{\alpha}{f} \,\sqrt{\frac{k}{2}}\,\int_{-\infty}^{\tau} d\tau'\, G_k (\tau,\tau')\,A_1(-k\tau')\,\partial_{\tau'} \delta\Phi(\tau').
\end{equation}
We can then insert this solution into the equation for the inflaton perturbation, \cref{eq:inf_eq}, obtaining the following integro-differential equation
\begin{align}\label{eq:deltaphi1}
    &\partial_\tau^2\delta\Phi(\tau)+
    2\frac{\partial_\tau a}{a}\,\partial_\tau\delta\Phi(\tau)+a^2\frac{d^2V(\bar\Phi)}{d\bar\Phi^2}\,\delta\Phi(\tau)\nonumber\\
    &=
    \frac{\alpha^2}{2\,f^2\,a^2}\int^{\tau}_{-\infty}   d\tau'\, \partial_{\tau'}\delta\Phi(\tau')\,\frac{\partial}{\partial \tau} \int \frac{d^3\bk}{(2\pi)^3}\, \text{Im}\left[ A_1(-k\tau)\, A_1^*(-k\tau') \right]\,\text{Re}\left[A_1(-k\tau)\,A_1^*(-k\tau')\right]\,.
\end{align}

This equation can be written in a more compact way as
\begin{equation}\label{eq:int_kernel}
    \partial_\tau^2\delta\Phi +2\frac{\partial_\tau a}{a}\,\partial_\tau\delta\Phi +a^2\frac{d^2V(\bar\Phi)}{d\bar\Phi^2}\,\delta\Phi
    =-\frac{\alpha^2}{f^2\,a^2\,\tau^4}\int^{\tau} d\tau' \,\partial_{\tau'}\delta\Phi(\tau')\,\mathcal{K}(\tau'/\tau).
\end{equation}
where we have defined the Kernel 
\begin{align}\label{eq:kernel}
    {\cal K}(z)\equiv \int \frac{x^3\,dx}{4\pi^2}\,\text{Im}\left[A_1(x)\,A_1'(x)\,A_1^*(x\,z)^2\right]\,,
\end{align}
and where we have performed the trivial integration on the angular coordinates of $\bk$ and changed the integration variable from $k$ to $x=-k\tau$. The integral in \cref{eq:kernel} is divergent in the UV. We cut it off at $x=x_+$, since modes with $x>x_+$ are still in their vacuum.

To find the general solution of \cref{eq:int_kernel}, we make the Ansatz $\delta\Phi=(-H_0 \tau)^{-\beta}$, with $\beta$ constant, for the inflaton perturbation. 
Plugging this Ansatz into \cref{eq:int_kernel} and using $a=(-H_0\tau)^{-1}$ we get an equation for $\beta$,
\begin{align}\label{eq:zeros1}
    \beta\,(\beta+3)
    +3\,\eta 
    +\frac{\alpha^2 H_0^2}{f^2}\,\beta\int_1^{\infty} dz \, z^{-\beta-1} \mathcal{K}(z)\equiv {\cal{I}}=0\,,
\end{align}
where the slow-roll parameter $\eta$ is defined in \cref{eq:def_sr} and where we have used the approximate Friedmann equation  $H_0^2\simeq V(\Phi)/3M_{\rm Pl}^2$. Solutions of \cref{eq:zeros1} where the real part of $\beta$ is positive correspond to unstable solutions. If, on the other hand, all solutions satisfy $\text{Re}(\beta)<0$, then the IBR solution with the given parameters $\xi_0$, $\mu_0$, $\eta$ and $\alpha\,H_0/f$ shows no instability.

\begin{figure}
\centering
\includegraphics[width=0.65\linewidth]{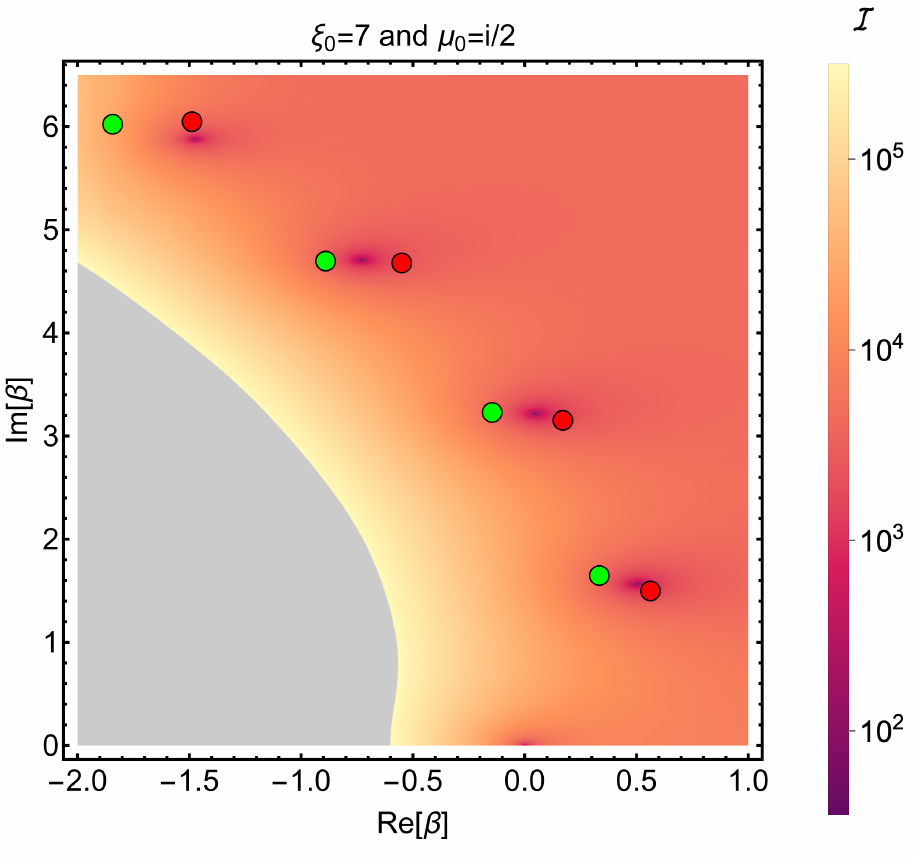}
\caption{Magnitude of ${\cal I}$, the left hand side of \cref{eq:zeros1}, in the massless case $\mu_0=i/2$ and for $\xi_0=7$. See the text under \cref{eq:zeros1} for details on how we treat the quantity $\alpha^2H_0^2/f^2$ appearing in ${\cal I}$. The gray area corresponds to ${\cal I}>10^{5.5}$. The purple dots correspond to the zeros of ${\cal I}$. Their positions are compared to those obtained in previous studies which use independent methods, in particular reference \cite{vonEckardstein:2023gwk} (red dots) and \cite{Peloso:2022ovc} (green dots).}
\label{fig:zeros_check}
\end{figure}

To test the validity of our formula, we first use \cref{eq:zeros1} to study the massless case ($\mu_0=i/2$) and check that it reproduces the results obtained previously in refs.~\cite{Peloso:2022ovc,vonEckardstein:2023gwk}. In particular, to compare with fig.~1 of ref.~\cite{vonEckardstein:2023gwk}, we use the empirical relation eq.~(2.55) in that article to trade the Hubble parameter for $\xi_0$ and we set $\xi_0=7$ and $\alpha/f=10^{2.5}/ M_{\rm Pl}$.\footnote{Note that the parameter $\beta$ in, e.g., Ref.~\cite{vonEckardstein:2023gwk} corresponds to $\alpha\,M_{\rm Pl}/f$ in our work.} In \cref{fig:zeros_check} we see that our results, obtained via a completely independent method, are in excellent agreement with those of previous studies.

We can now move on to solving \cref{eq:zeros1} for the case of massive vectors we are interested in. First, we can use \cref{eq:back_const} to trade $H_0$ for other parameters in the theory, obtaining an equation that we can write as
\begin{align}\label{eq:zeros2}
    \left(\sqrt{\frac{\epsilon}{2}}\,\frac{\alpha\,M_P}{f}-\xi_0\right)\, \int_1^{\infty} dz \, z^{-\beta-1} \mathcal{K}(z)\simeq -e^{2\pi\,(\xi_0-\mu_0)}\,\left((\beta+3)
    +3\,\frac{\eta}{\beta} \right)\frac{0.8}{48\pi^2}\frac{\xi_0^{7/2}}{(\xi_0-\mu_0)^{1/6}}\,.
\end{align}
This equation depends on five parameters: the two slow-roll parameters $\epsilon$ and $\eta$, the rescaled coupling parameter $\alpha\,M_{\rm Pl}/f$, and the two dimensionless parameters $\xi_0$ and $\mu_0$ defined in \cref{eq:xiandmu}.

Note that there are some constraints on the values of these parameters. In particular, from \cref{eq:back_const} we see that $\alpha\,M_{\rm Pl}\sqrt{2\epsilon}>2\,\xi_0\,f$. Moreover, the exponential dependence on $\pi(\xi_0-\mu_0)$ of the right hand side of \cref{eq:back_const} implies that $\xi_0-\mu_0$ cannot take values much larger than ${\cal O}(1)$ or so. Finally, {\it if} the system is in the strong backreaction regime when cosmological scales left the horizon, then we can match the predicted amplitude of the spectrum of scalar perturbations, \cref{eq:pz_approx}, to its observed value, which leads to $\xi_0\simeq 300$. Of course, the system might also enter the strong backreaction regime after the cosmological scales have left the horizon, in which case $\xi_0$ becomes a free parameter again. Nevertheless, we will take this as an indication that a ``natural'' value for $\xi_0$ is $\xi_0={\cal O}(100)$.

In principle, scanning a five-dimensional parameter space would be a challenge, but things are simplified by the fact that, for the range of $\xi_0$ and $\mu_0$ we have explored (${\cal O}(1)\lesssim \xi_0-\mu_0\ll \xi_0\lesssim {\cal O}(10^3)$), numerical evaluations show that the integral $\int_1^{\infty} dz \, z^{-\beta-1} \mathcal{K}(z)$ takes values that are typically much larger than the quantity appearing on the right hand side of \cref{eq:zeros2}. 
Therefore, as long as $\sqrt{\frac{\epsilon}{2}}\,\frac{\alpha\,M_P}{f}-\xi_0$ is not very small, the solutions of \cref{eq:zeros2} are very close to the solutions of the much simpler equation
\begin{align}\label{eq:zeros3}
   \int_1^{\infty} dz \, z^{-\beta-1} \mathcal{K}(z)\simeq 0\,,
\end{align}
which depends only on $\xi_0$ and $\mu_0$.  We also note that the condition ``$\sqrt{\frac{\epsilon}{2}}\,\frac{\alpha\,M_P}{f}-\xi_0$ is not very small'' quoted above corresponds precisely to the strong backreaction condition, as this quantity exactly vanishes in the absence of backreaction.

\begin{figure}
\centering
\begin{subfigure}{.5\textwidth}
  \centering
  \includegraphics[width=1\linewidth]{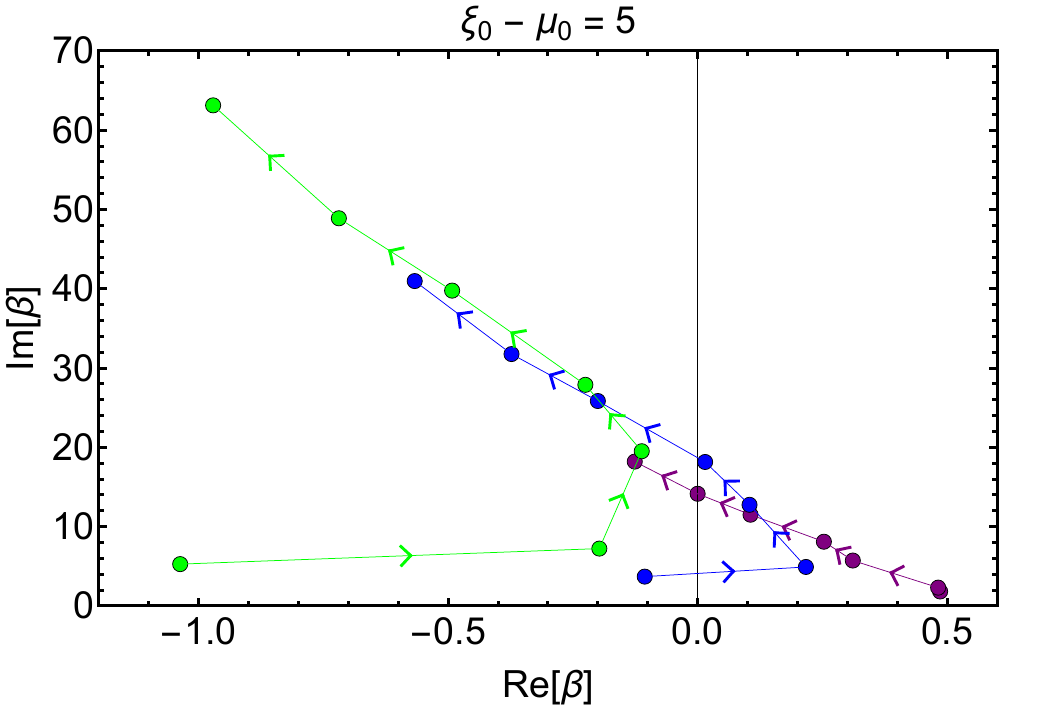}
\end{subfigure}%
\begin{subfigure}{.5\textwidth}
  \centering
  \includegraphics[width=1\linewidth]{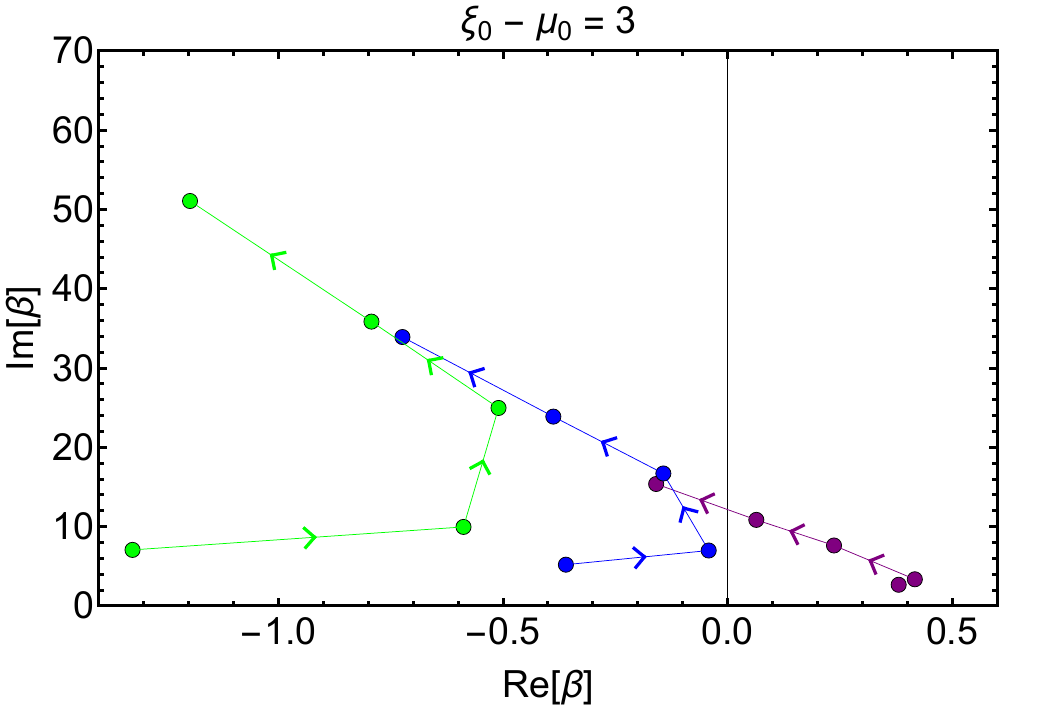}
\end{subfigure}
\caption{Position of the first three zeros of ${\cal I}$, the left hand side of \cref{eq:zeros1}, using \cref{eq:zeros3} for different values of $\xi_0$ and $\mu_0$. For $\xi_0-\mu_0=5$ (left) the arrows follow $\xi_0=6 \rightarrow 10 \rightarrow 50 \rightarrow 100 \rightarrow 200 \rightarrow 300 \rightarrow 500$. For $\xi_0-\mu_0=3$ (right) the arrows follow $\xi_0=6 \rightarrow 10 \rightarrow 50 \rightarrow 100 \rightarrow 200$. Note that the first two purple dots for $\xi_0=6$ and $\xi_0=10$ are almost overlapping in both cases.}
\label{fig:First_zeros}
\end{figure}

We solve \cref{eq:zeros3} for two fixed values of $\xi_0-\mu_0=3\,,5$. In \cref{fig:First_zeros}, we show the position of the first three zeros (i.e., the three solutions to \cref{eq:zeros3} with the largest value of $\text{Re}(\beta)$) as we increase the value of $\xi_0$.

These plots, which represent the main result of our work, show the presence of unstable solutions with $\text{Re}(\beta)>0$ for small values of $\xi_0$. However, as we increase $\xi_0$, all zeros shift towards the region where $\text{Re}(\beta)<0$, indicating that, as we increase the mass of the vector field for fixed values of the Hubble parameter and of $\xi_0-\mu_0$, the system moves towards stability.  In particular, we observe that for $\xi_0-\mu_0=5$, all solutions become stable for $\xi_0 \gtrsim 300$, while for $\xi_0-\mu_0=3$ they are stable for $\xi_0 \gtrsim 120$. 

\begin{figure}
\centering
\begin{subfigure}{.5\textwidth}
  \centering
  \includegraphics[width=1\linewidth]{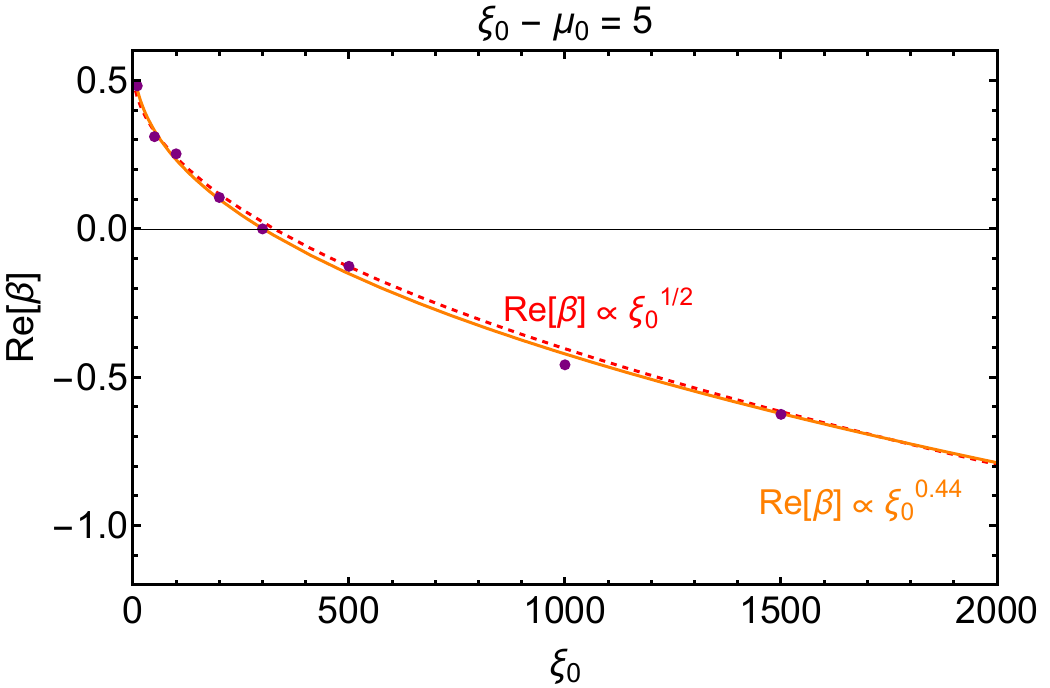}
\end{subfigure}%
\begin{subfigure}{.5\textwidth}
  \centering
  \includegraphics[width=1\linewidth]{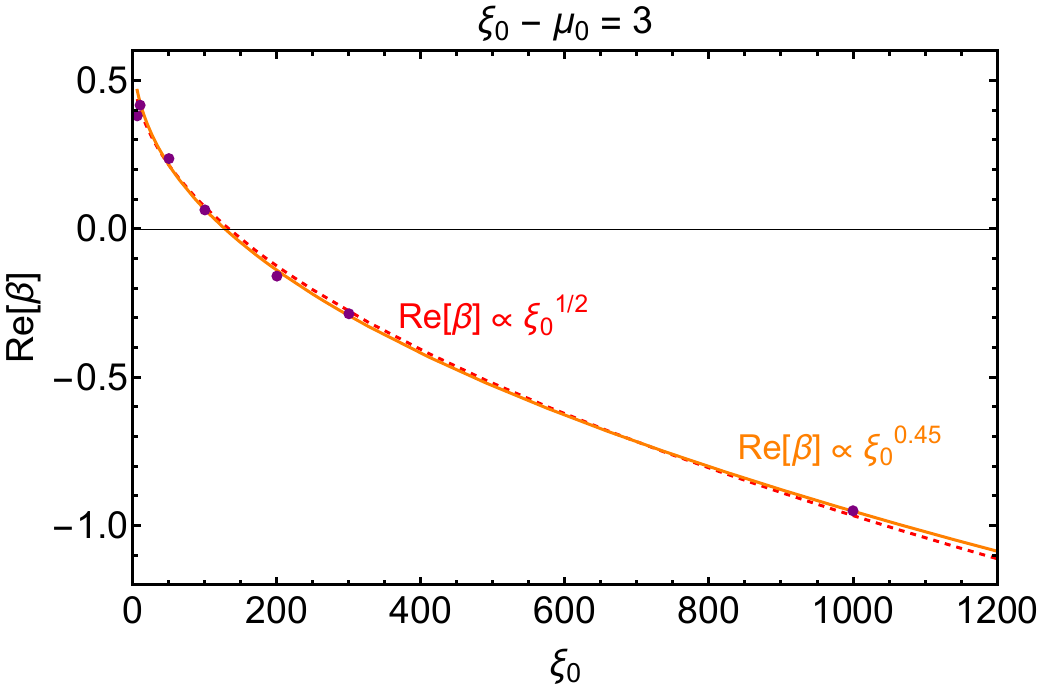}
\end{subfigure}
\caption{Real part of the zero of \cref{eq:zeros3} with largest real part  (purple dots in \cref{fig:First_zeros}) as a function of $\xi_0$ for $\xi_0-\mu_0=5$ (left) and for $\xi_0-\mu_0=3$ (right).}
\label{fig:First_zero_fit}
\end{figure}

In \cref{fig:First_zero_fit} we plot the value of the real part of the first zero of \cref{eq:zeros3} as a function of $\xi_0$. For the values of $\xi_0$ and $\mu_0$ we consider, we find a good power-law fit 
\begin{align}
    \text{Re}(\beta)\approx \kappa_1-\kappa_2\,\sqrt{\xi_0}\,,
\end{align}
with $\kappa_1$ and $\kappa_2$ positive functions of $\xi_0-\mu_0$ only. 

\section{Conclusions and Future Directions}
\label{sec:conclusions}

In this work we have studied, to our knowledge for the first time, a model of axion inflation in which the inflaton is coupled through a Chern--Simons interaction to a {\it massive} vector field. Although the vector field has been modeled with a Proca action, it can also be seen as the low-energy limit of a gauge field that acquires its mass through the Higgs mechanism, with the Higgs field heavy enough to decouple from the dynamics of the system.

We have shown that if the vector is massive enough, the transition to the strong backreaction regime occurs smoothly and the evolution of the system shows no instabilities. In other words, the system does not exhibit significant memory effects and the vector field can be integrated out, yielding an effective equation of motion for the homogeneous inflaton that is local in time and can be solved as an ordinary differential equation. We have referred to this as the ``instant backreaction'' (IBR) regime. Therefore, this is a modification of the model of axion inflation where the late time dynamics can be reliably calculated. This behavior should be contrasted with the case of massless vectors, where memory effects render the dynamics so intricate that, despite extensive work over the past decade, the evolution of the system in that case is still not fully understood.

Our results raise a number of interesting directions for future work. For certain choices of the model parameters, the system transitions to strong backreaction late in inflation, well after the cosmological scales have crossed the horizon. It will be interesting to investigate how the phenomenology associated with the final stages of inflation -- such as the production of gravitational waves at interferometer scales or the formation of primordial black holes -- is modified relative to the well-studied case of massless photons.

A much less explored possibility is that the last $\sim 60$ e-folds of inflation took place entirely in the strong backreaction regime~\cite{Anber:2009ua}, leading to a slow roll mechanism that does not rely on a flat potential. Establishing the phenomenologicalle viability of this scenario should be a priority. This will require a more reliable determination of the scalar power spectrum than the estimate presented in \cref{subsec:Pspectrum}, together with a computation of the tensor spectrum. It will also be important to evaluate the level of primordial non-Gaussianity and the scalar spectral index, both of which are tightly constrained by observations. More broadly, because the connection between observables such as the amplitude and tilt of the scalar power spectrum and the underlying model parameters differs substantially from that in conventional slow-roll inflation, this framework opens up a wide range of new possibilities for inflationary model building.

\section*{Acknowledgments}
\label{sec:acknowledgments}

We thank Mohamed Anber and Juan Pablo Beltr\'an Almeida for helping to 
launch this work and for making important contributions during its early stages. We also thank Kai Schmitz, Richard von Eckardstein and especially Sasha Sobol for sharing part of their data. The work of L.S. is partially supported by the US-NSF grant PHY-2412570.

\appendix

\section{WKB Approximation for the Mode Functions}
\label{app:wkb}

In this appendix we present formulae for the mode functions in the WKB approximation. We remind the reader that we use $x_\pm$ (defined in \cref{eq:xpm}) to denote the zeros of the effective frequency for the mode functions of the positive helicity modes of our massive vector,
\begin{align}\label{eq:defomega}
    \omega(x)\equiv \sqrt{1-\frac{2\,\xi_0}{x}+\frac{\mu_0^2}{x^2}}
    \,,
\end{align}
see \cref{Apm}. In this appendix, as well as in most of our paper, we will work in the regime $\mu_0\gg 1$. As a consequence, to simplify our expressions, we will write $\mu_0$ instead of $\sqrt{\mu_0^2+1/4}$.

For $x\to+\infty$ we must keep only the positive frequency mode, $\sqrt{2\,k}\,A_+(x)\equiv A_1(x)\approx e^{i\,x}$. To match this behavior,  we  choose the integration constants for the WKB-approximated mode function (that is, taking $|\partial_x\omega|/\omega^2 \ll 1$) for $x>x_+$ as   
\begin{align}\label{eq:wkb_large}
    A^{\rm WKB}_1(x>x_+)= \frac{e^{-i\pi/4+i\int_{x_+}^x\omega(y)\,dy}}{\omega(x)^{1/2}}\,,
\end{align}
where we have fixed the arbitrary phase to $e^{-i\pi/4}$ to simplify the later formulae. The integral in the phase (as well as all the other integrals in this appendix) can be computed explicitly in terms of elementary functions, but these expressions are not terribly illuminating and we will not report them here.

In the region $x_-<x<x_+$ the vector field mode functions have an imaginary frequency and, in the WKB approximation, will be a linear combination of $(-1+2\,\xi_0/x-\mu_0^2/x^2)^{-1/4}\times e^{\pm\int_{x_-}^{x}\sqrt{-1+2\,\xi_0/y-\mu_0^2/y^2}\,dy}$. The coefficients of the linear combination can be obtained by solving exactly the equation obtained by linearizing \cref{Apm} around $x=x_+$ and matching the solution to the  expressions in WKB approximation for $x>x_+$ and $x<x_+$. This procedure yields
\begin{align}\label{eq:wkb_int} 
A^{\rm WKB}_1(x_-<x<x_+)=&\frac{1}{\sqrt{\kappa(x)}}\left[-i\,e^{\pi(\xi_0-\mu_0)-\int_{x_-}^x \kappa(y)\,dy}+\frac12\,e^{-\pi(\xi_0-\mu_0)+\int_{x_-}^x\kappa(y)\,dy}\right]\,,\nonumber\\
&\qquad\qquad\qquad \kappa(x)\equiv \sqrt{-1+\frac{2\,\xi_0}{x}-\frac{\mu_0^2}{x^2}}\,,
\end{align}
where we have used the result $\int_{x_-}^{x_+}\kappa(y)\,dy=\pi(\xi_0-\mu_0)\,$. This expression shows that as $x\lesssim x_+$ the mode functions quickly attain an amplitude of the order of $e^{\pi(\xi_0-\mu_0)}$.

Finally, using the same procedure around $x_-$, we write the mode functions for $x<x_-$ as
\begin{align}\label{eq:wkb_small}
A^{\rm WKB}_1(x<x_-)=-\frac{2\,i}{\sqrt{\omega(x)}}&\Big[e^{\pi(\xi_0-\mu_0)}\,\cos\left(\int_{x_-}^x\omega(y)\,dy-\frac{\pi}{4}\right)\nonumber\\
&-\frac{i}{4}\,e^{-\pi(\xi_0-\mu_0)}\,\sin\left(\int_{x_-}^x\omega(y)\,dy-\frac{\pi}{4}\right)\Big]\,,
\end{align}
where the term in the second line is exponentially suppressed and can be neglected.

\section{Estimating the Backreaction Integral}
\label{app:backreact}

In this appendix we derive the approximate expression \cref{eq:back_int_approx} of the backreaction integral appearing in equation~(\ref{eq:change_kx}).

Our starting point is the WKB approximation of the vector's mode function, eqs.~(\ref{eq:wkb_int}) and~(\ref{eq:wkb_small}), from which we obtain, after neglecting the exponentially suppressed components,
\begin{align}\label{eq:aplusprime_wkb}
\frac{d\left|A^{\rm WKB}_1(x)\right|^2}{dx}\approx
e^{2\pi(\xi_0-\mu_0)}\times\begin{cases}
\begin{array}{ll}
-4\,\cos\left(2\int^{x_-}_x\omega(y)\,dy\right)\,, & x<x_-\\
-2\,\exp\left(-2\int_{x_-}^{x} \kappa(y)\,dy\right)\,, & x_-<x<x_+
\end{array}
\end{cases}
\,,
\end{align}
where $\omega(x)$ and $\kappa(x)$ are defined in eqs.~(\ref{eq:defomega}) and~(\ref{eq:wkb_int}). We cut off the integral at $x=x_+$ as for larger values of $x$ the modes of the gauge field are in their vacuum.

Inspection of the behavior of the function in eq.~(\ref{eq:aplusprime_wkb}) shows that the integral is dominated by the region around $x=x_-$. Expanding the integrand around that point we have, for instance, 
\begin{align}
\int_{x_-}^x\omega(y)\,dy&=\int_{x}^{x_-}\sqrt{\frac{(x_--y)(x_+-y)}{y^2}}\,dy\approx\int_{x}^{x_-}\sqrt{\frac{(x_--y)(x_+-x_-)}{x_-^2}}\,dy\nonumber\\
&=\frac23\left(x_--x\right)^{3/2}\frac{\sqrt{x_+-x_-}}{x_-}\,,
\end{align}
so the part of the integral we need to compute is approximately given by
\begin{align}
&\int_0^{x_-} x^3\,dx\,\cos\left(\frac43\left(x_--x\right)^{3/2}\frac{\sqrt{x_+-x_-}}{x_-}\right)\nonumber\\
&=\frac{x_-^{11/3}}{[6(x_+-x_-)]^{1/3}}\int_0^{\frac43\sqrt{x_-(x_+-x_-)}} \left[1-\left(\frac{3\,z}{4\,\sqrt{x_-}\sqrt{x_+-x_-}}\right)^{2/3}\right]^3\cos\left(z\right)\,\frac{dz}{z^{1/3}}\,,
\end{align}
where we have changed integration variable to $z= \frac43\left(x_--x\right)^{3/2}\frac{\sqrt{x_+-x_-}}{x_-}$. Next, in the 
regime $1\lesssim \xi_0-\mu_0\ll \xi_0$ we are interested in, we have $\sqrt{x_-}\sqrt{x_+-x_-}\gg 1$, so that we can approximate the integral with
\begin{align}
&\frac{x_-^{11/3}}{[6(x_+-x_-)]^{1/3}}\int_0^\infty \cos\left(z\right)\,\frac{dz}{z^{1/3}}=\frac{\Gamma(2/3)}{2\times 6^{1/3}}\,\frac{x_-^{11/3}}{(x_+-x_-)^{1/3}}\,.
\end{align}
With similar manipulations we obtain
\begin{align}
    \int_{x_-}^{x_+} x^3\,dx\,\exp\left(-2\int_{x_-}^{x} \kappa(y)\,dy\right)\simeq \frac{\Gamma(2/3)}{6^{1/3}}\frac{x_-^{11/3}}{(x_+-x_-)^{1/3}}
    \,.
\end{align}

By collecting these results, we can finally write
\begin{align}\label{eq:back_wkb_almostfinal}
\int dx \, x^3 \, \frac{d\left|A^{\rm WKB}_1(x)\right|^2}{d x} \simeq -4\,\frac{\Gamma(2/3)}{6^{1/3}}\,e^{2\pi(\xi_0-\mu_0)}\,\frac{x_-^{11/3}}{(x_+-x_-)^{1/3}}\,.
\end{align}
In particular, for $\xi_0-\mu_0>1$ but not too large and $\xi_0\to\infty$, one sees that the right hand side of eq.~(\ref{eq:back_wkb_almostfinal}) scales as $e^{2\pi\,(\xi_0-\mu_0)}\xi_0^{7/2}/(\xi_0-\mu_0)^{1/6}$. We have checked the validity of our estimate for a range of values of $\xi_0$ and $\mu_0$ satisfying $1\lesssim \xi_0-\mu_0\ll \xi_0$ and we have found that it works very well except for an overall ${\cal O}(1)$ factor, which we attribute to the fact that the WKB approximation fails by ${\cal O}(1)$ around $x=x_-$. Fixing ``by hand'' that overall factor, we have found that the expression
\begin{align}
\int dx \, x^3 \, \frac{d\left|A_1(x)\right|^2}{d x} \simeq -0.8\times e^{2\pi(\xi_0-\mu_0)}\,\frac{\xi_0^{7/2}}{(\xi_0-\mu_0)^{1/6}}\,,
\end{align}
is accurate at the $\sim 50\%$  level for $2\le\xi_0-\mu_0<10$ and $\xi_0>50$.

\section{Estimating the Power Spectrum of the Curvature Perturbation}
\label{app:powerspectrum}

In order to compute the power spectrum ${\cal P}_\zeta$ we need to Fourier transform \cref{eq:zeta_int} and take its two-point function, leading to a multiple integral that can be written in a compact way as
\begin{align}
    \langle\zeta(\tau,\,\bk_1)\,\zeta(\tau,\,\bk_2)\rangle&=\frac{1}{(2\pi\,\xi_0)^2}\int_{-\infty}^\tau\frac{d\tau_1}{\tau_1}\int_{-\infty}^\tau\frac{d\tau_2}{\tau_2}\int\frac{d^3\bp_1\,d^3\bp_2}{(2\pi)^3}\nonumber\\
    &\times\frac{\langle({\bf E}(\tau_1\,,\bp_1)\cdot{\bf B}(\tau_1\,,-\bp_1-\bk_1))\,({\bf E}(\tau_1\,,\bp_2)\cdot{\bf B}(\tau_2\,,-\bp_2-\bk_2))\rangle_{\rm {conn}}}{\langle{\bf E}\cdot{\bf B}\rangle(\tau_1)\,\langle{\bf E}\cdot{\bf B}\rangle(\tau_2)}
    \,,
\end{align}
where the notation $\langle\ldots\rangle_{\rm {conn}}$ indicates that when performing the Wick reduction of the expectation value of the four operators into sums of products of two-point functions, one should neglect the disconnected diagrams (tadpoles) that are canceled by the expectation value $\langle{\bf E}\cdot{\bf B}\rangle$ appearing in the numerator of \cref{eq:zeta_int}.

With brute force calculation we can write
\begin{align}
&\int d^3\bp_1\,d^3\bp_2\langle({\bf E}(\tau_1\,,\bp_1)\cdot{\bf B}(\tau_1\,,-\bp_1-\bk_1))\,({\bf E}(\tau_1\,,\bp_2)\cdot{\bf B}(\tau_2\,,-\bp_2-\bk_2))\rangle_{\rm {conn}}\nonumber\\
&=\delta(\bk_1+\bk_2)\int d^3\bp_1\,\left|\epsilon_+(\bp_1)\cdot\epsilon_+(-\bp_1-\bk_1)\right|^2\nonumber\\
&\times\Big[\partial_{\tau_1}A_+(\tau_1\,,p_1)\,\partial_{\tau_2}A_+(\tau_2\,,p_1)^*\, |\bp_1+\bk_1|^2\,A_+(\tau_1\,,|\bp_1+\bk_1|)\,A_+(\tau_2\,,|\bp_1+\bk_1|)^*\nonumber\\
&+\partial_{\tau_1}A_+(\tau_1\,,p_1)\,|\bp_1|\,A_+(\tau_2\,,p_1)^*\,|\bp_1+\bk_1|\,A_+(\tau_1\,,|\bp_1+\bk_1|)\,\partial_{\tau_2}A_+(\tau_2\,,|\bp_1+\bk_1|)^*\Big]\,,
\end{align}
which can be further simplified by writing it in terms of the function $A_1(x)$ defined in \cref{Apmsol}:
\begin{align}
&\int d^3\bp_1\,d^3\bp_2\,\langle({\bf E}(\tau_1\,,\bp_1)\cdot{\bf B}(\tau_1\,,-\bp_1-\bk_1))\,({\bf E}(\tau_1\,,\bp_2)\cdot{\bf B}(\tau_2\,,-\bp_2-\bk_2))\rangle_{\rm {conn}}\nonumber\\
&=\frac{\delta(\bk_1+\bk_2)}{16}\int d^3\bp_1\,\left(1+\frac{\bp_1\cdot(\bp_1+\bk_1)}{p_1\,|\bp_1+\bk_1|}\right)^2\,p_1\,|\bp_1+\bk_1|\,A_1'(-p_1\tau_1)\,A_1(-|\bp_1+\bk_1|\tau_1))\nonumber\\
&\times\Big[A_1'(-p_1\tau_2)^*\,A_1(-|\bp_1+\bk_1|\tau_2)^*+A_1(-p_1\tau_2)^*\,A_1'(-|\bp_1+\bk_1|\tau_2)^*\Big]\,.
\end{align}

To proceed we note that \cref{eq:eom-2.4,eq:kg_backr,eq:change_kx} show that
\begin{align}
    \langle{\bf E}\cdot{\bf B}\rangle(\tau)=-\frac{1}{8\pi^2\,\tau^4}\int dx\,x^3\,\frac{d\left|A_1(x)\right|^2}{dx}\equiv \frac{{\cal J}}{8\pi^2\tau^4}\,,
\end{align}
with ${\cal J}$ a constant depending only on $\xi_0$ and $\mu_0$.  Defining the power spectrum ${\cal P}_\zeta$ as
\begin{align}
    \langle\zeta(\tau,\,\bk_1)\,\zeta(\tau,\,\bk_2)\rangle\equiv 2\pi^2\,\frac{\delta(\bk_1+\bk_2)}{k_1^3}\,{\cal P}_\zeta(\tau,\,k_1)
\end{align}
we finally obtain the expression 
\begin{align}
    &{\cal P}_\zeta(\tau,\,k_1)=\frac{1}{2^4\,\pi^3\,\xi_0^2\,{\cal J}^2}\int^{k_1\tau}_{-\infty} d\tau_1\,\tau_1^3\,\int^{k_1\tau}_{-\infty} d\tau_2\,\tau_2^3\int d^3\bp_1\,\left(1+\frac{\bp_1\cdot(\bp_1+\hat\bk_1)}{p_1\,|\bp_1+\hat\bk_1|}\right)^2\,\nonumber\\
    &\qquad\qquad\qquad\times p_1\,|\bp_1+\hat\bk_1|\,A_1'(-p_1\tau_1)\,A_1(-|\bp_1+\hat\bk_1|\tau_1)\nonumber\\
    &\qquad\qquad\times\Big[A_1'(-p_1\tau_2)^*\,A_1(-|\bp_1+\hat\bk_1|\tau_2)^*+A_1(-p_1\tau_2)^*\,A_1'(-|\bp_1+\hat\bk_1|\tau_2)^*\Big]
    \,,
\end{align}
where we have redefined the integration variables $\tau_1$, $\tau_2$ and $\bp_1$ by rescaling them by appropriate powers of $k_1$ and where $\hat{\bk}_1$ denotes the unit vector directed along $\bk_1$. Since we evaluate the power spectrum at the end of inflation, $\tau=-1/H$, and for superhorizon modes, $k_1\ll H$, we will set the upper limits of integration in $d\tau_1$ and $d\tau_2$ to zero. In this limit the power spectrum is independent of $k_1$, showing that the resulting spectrum of scalar perturbations is scale invariant in our regime of approximations. 

\begin{figure}{h}
\centering
  \includegraphics[width=.5\linewidth]{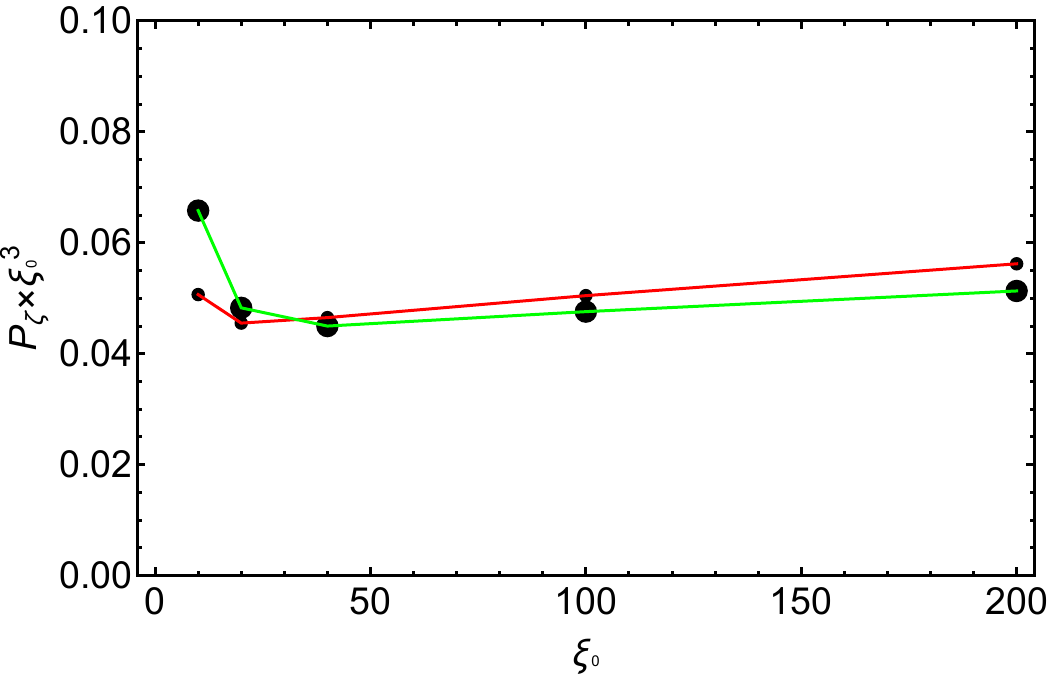}
\caption{Power spectrum ${\cal P}_\zeta$ as obtained from \cref{eq:final_pzeta}, multiplied by $\xi_0^3$, for various values of $\xi_0$ and fixed $\xi_0-\mu_0=3$ (smaller dots, red line) and $\xi_0-\mu_0=5$ (larger dots, green line). }
\label{fig:powerspectrum}
\end{figure}

It is convenient to define the function
\begin{align}
    F(k,\,q)\equiv \int_{-\infty}^0 d\tau\,\tau{}^3\left(A_1'(-k\tau)\,A_1(-q\tau)+A_1'(-q\tau)\,A_1(-k\tau)\right)\,,
\end{align}
so that the expression of ${\cal P}_\zeta(\tau,\,k_1)$ can be written in a much more compact way as
\begin{align}
    &{\cal P}_\zeta=\frac{1}{2^5\pi^3\xi_0^2\,{\cal J}^2} \int d^3\bp_1\,\left(1+\frac{\bp_1\cdot(\bp_1+\hat\bk_1)}{p_1\,|\bp_1+\hat\bk_1|}\right)^2\,\frac{|\bp_1+\hat\bk_1|}{p_1^7}\,
    \left|F\left(1,\,\frac{|\bp_1+\hat\bk_1|}{p_1}\right)\right|^2
    \,,
\end{align}
where we have used the scaling $F(k,\,q)=F(1,\,q/k)/k^4$. In order to evaluate numerically the integrals in this last equation, it proves convenient to direct $\hat\bk_1$ along the $z$-axis and redefine the integration variables to 
\begin{align}
    p_1=X+Y\,,\qquad\qquad |\bp_1+\hat\bk_1|=X-Y\,,
\end{align}
(for details see Appendix A in \cite{Garcia-Bellido:2023ser}). The final expression for the scalar power spectrum is
\begin{align}\label{eq:final_pzeta}
    &{\cal P}_\zeta=\frac{1}{2^5\pi^2\xi_0^2\,{\cal J}^2} \int_{1/2}^\infty dX\,\int_{-1/2}^{1/2}\,dY \, \frac{(4\,X^2-1)^2}{(X+Y)^8}\,\left|F\left(1,\,\frac{X-Y}{X+Y}\right)\right|^2\,.
\end{align}

We show in \cref{fig:powerspectrum} the resulting power spectrum ${\cal P}_\zeta$, multiplied by $\xi_0^3$, for a set of values of $\xi_0$ ranging between $10$ and $200$, and for $\xi_0-\mu_0=3,\,5$. The plot indicates that the power spectrum is well approximated by \cref{eq:pz_approx} in the main text.

\bibliographystyle{utphys}
\bibliography{biblio}  

\nocite{*}

\end{document}